\documentclass[longauth]{aa}

\usepackage{graphicx}
\usepackage{txfonts}
\usepackage[colorlinks=true,linkcolor=blue,citecolor=blue,breaklinks]{hyperref}

\usepackage[utf8]{inputenc}
\usepackage{caption}
\usepackage{subcaption}
\usepackage{multirow}
\usepackage{color}
\usepackage{amsmath}
\usepackage{etoolbox}
\usepackage{booktabs}
\usepackage{orcidlink}
\usepackage[separate-uncertainty=true, multi-part-units=single,range-phrase=--, range-units=single]{siunitx}
\usepackage{breqn}
\usepackage[switch]{lineno}

\newcommand{\eg}{e.g.\@\xspace}
\newcommand{\ie}{i.e.\@\xspace}
\newcommand{\CitepInPrep}[1]{(#1 et al.\@ in prep.)}
\newcommand{\CiteInPrep}[1]{#1 et al.\@ (in prep.)}
\newcommand{\resp}{resp.\@\xspace}
\newcommand{\Eq}{Eq.\@\xspace}
\newcommand{\Sect}{Sect.\@\xspace}
\newcommand{\Fig}{Fig.\@\xspace}

\DeclareSIUnit\arcsec{as}
\DeclareSIUnit\arcmin{am}
\DeclareSIUnit\angstrom{\text{\normalfont\AA}}
\DeclareSIUnit\pixel{pixel}
\DeclareSIUnit\mag{mag}
\DeclareSIUnit{\inch}{in}
\DeclareSIUnit{\electron}{e\textsuperscript{-}}
\DeclareSIUnit{\ADU}{ADU}
\DeclareSIUnit\byte{byte}

\begin{document}

   \title{ZTF SN~Ia DR2: Toward cosmology-grade ZTF supernova light curves using scene modeling photometry}

   \subtitle{}

   \author{L. Lacroix
            \inst{\ref{lpnhe}, \ref{okc_physics},\ref{ip2i}}\thanks{\email{llacroix@ip2i.in2p3.fr}}\orcidlink{0000-0003-0629-5746}
            \and
            N. Regnault\inst{\ref{lpnhe}}\thanks{\email{nicolas.regnault@lpnhe.in2p3.fr}}\orcidlink{0000-0001-7029-7901}
            \and
            T. de Jaeger\inst{\ref{lpnhe}}\orcidlink{0000-0001-6069-1139}
            \and
            M. Le Jeune\inst{\ref{apc}}\orcidlink{0000-0002-1008-3394}
            \and
            M. Betoule\inst{\ref{lpnhe}}\orcidlink{0000-0003-0804-836X}
            \and
            J.-M. Colley\inst{\ref{lpnhe}}\orcidlink{0009-0008-8650-8775}
            \and
            M. Bernard\inst{\ref{lpnhe}}\orcidlink{0000-0001-7586-7133}
            \and
            M. Rigault\inst{\ref{ip2i}}\orcidlink{0000-0002-8121-2560}
            \and
            M. Smith\inst{\ref{lancaster}}\orcidlink{0000-0002-3321-1432}
            \and
            A. Goobar\inst{\ref{okc_physics}}\orcidlink{0000-0002-4163-4996}
            \and
            K. Maguire\inst{\ref{trinity}}\orcidlink{0000-0002-9770-3508}
            \and
            G. Dimitriadis\inst{\ref{lancaster}}\orcidlink{0000-0001-9494-179X}
            \and
            J. Nordin\inst{\ref{humboldt}}\orcidlink{0000-0001-8342-6274}
            \and
            J. Johansson\inst{\ref{okc_physics}}\orcidlink{0000-0001-5975-290X}
            \and
            M. Aubert\inst{\ref{lpca}}\orcidlink{0009-0002-7667-8814}
            \and
            C. Barjou-Delayre \inst{\ref{lpca}}\orcidlink{0009-0000-8510-8982}
            \and
            E. C. Bellm\inst{\ref{dirac}}\orcidlink{0000-0001-8018-5348}
            \and
            S. Bongard\inst{\ref{lpnhe}}\orcidlink{0000-0002-3399-4588}
            \and
            U. Burgaz\inst{\ref{trinity}}\orcidlink{0000-0003-0126-3999}
            \and
            B. Carreres\inst{\ref{duke}}\orcidlink{0000-0002-7234-844X}
            \and
            D. Fouchez\inst{\ref{cppm}}\orcidlink{0000-0002-7496-3796}
            \and
            F. Feinstein\inst{\ref{cppm}}\orcidlink{0000-0001-5548-3466}
            \and
            L. Galbany\inst{\ref{ice_csic},\ref{ieec}}\orcidlink{0000-0002-1296-6887}
            \and
            M. Ginolin\inst{\ref{ip2i}}\orcidlink{0009-0004-5311-9301}
            \and
            M. Graham\inst{\ref{caltech_phys}}\orcidlink{0000-0002-3168-0139}
            \and
            D. Kuhn\inst{\ref{lpnhe}}\orcidlink{0009-0005-8110-397X}
            \and
            R. R. Laher\inst{\ref{ipac}}\orcidlink{0000-0003-2451-5482}
            \and
            T. E. Müller-Bravo\inst{\ref{trinity},\ref{icen_chile}}\orcidlink{0000-0003-3939-7167}
            \and
            J. Neveu\inst{\ref{lpnhe},\ref{ijclab}}\orcidlink{0000-0002-6966-5946}
            \and
            M. Osman\inst{\ref{lpnhe}}\orcidlink{0009-0000-6101-6725}
            \and
            B. Popovic\inst{\ref{ip2i}}\orcidlink{0000-0002-8012-6978}
            \and
            B. Racine\inst{\ref{cppm}}\orcidlink{0000-0001-8861-3052}
            \and
            P. Rosnet\inst{\ref{lpca}}\orcidlink{0000-0002-6099-7565}
            \and
            D. Rosselli\inst{\ref{cppm}}\orcidlink{0000-0001-6839-1421}
            \and
            R. Smith\inst{\ref{caltech_obs}}\orcidlink{0000-0001-7062-9726}
            \and
            J. Sollerman\inst{\ref{okc_astro}}\orcidlink{0000-0003-1546-6615}
            \and
            J. H. Terwel\inst{\ref{trinity}}\orcidlink{0000-0001-9834-3439}
            \and
            A. Townsend\inst{\ref{humboldt}}\orcidlink{0000-0001-6343-3362}
            \and
            A. Wold\inst{\ref{ipac}}\orcidlink{0000-0002-9998-6732}
        }

        \institute{
        Sorbonne Université, Université Paris Cité, CNRS, Laboratoire de Physique Nucléaire et de Hautes Energies, 4 Place Jussieu, 75252 Paris, France
        \label{lpnhe}
        \and
        The Oskar Klein Centre, Department of Physics, Stockholm University, Albanova University Center, Stockholm, SE-106 91, Sweden
        \label{okc_physics}
        \and
        Universite Claude Bernard Lyon 1, CNRS, IP2I Lyon / IN2P3, IMR 5822, F-69622 Villeurbanne, France \label{ip2i}
        \and
        Université de Paris, CNRS, Astroparticule et Cosmologie, F-75013 Paris, France
        \label{apc}
        \and
        Department of Physics, Lancaster University, Lancaster LA1 4YB, UK
        \label{lancaster}
        \and
        School of Physics, Trinity College Dublin, The University of Dublin, Dublin 2, Ireland
        \label{trinity}
        \and
        Institut für Physik, Humboldt-Universität zu Berlin, Newtonstr. 15, 12489 Berlin, Germany
        \label{humboldt}
        \and
        Université Clermont Auvergne, CNRS/IN2P3, LPCA, F-63000 Clermont-Ferrand, France
        \label{lpca}
        \and
        DIRAC Institute, Department of Astronomy, University of Washington, 3910 15th Avenue NE, Seattle, WA 98195, USA
        \label{dirac}
        \and
        Department of Physics, Duke University, Durham, NC 27708, USA
        \label{duke} 
        \and
        Aix Marseille Univ, CNRS/IN2P3, CPPM, Marseille, France
        \label{cppm}
        \and
        Institute of Space Sciences (ICE-CSIC), Campus UAB, Carrer de Can Magrans, s/n, E-08193 Barcelona, Spain
        \label{ice_csic}
        \and
        Institut d'Estudis Espacials de Catalunya (IEEC), 08860 Castelldefels (Barcelona), Spain
        \label{ieec}
        \and
        Division of Physics, Mathematics and Astronomy, California Institute of Technology, Pasadena, CA 91125, USA
        \label{caltech_phys}
        \and
        IPAC, California Institute of Technology, 1200 E. California Blvd, Pasadena, CA 91125, USA
        \label{ipac}
        \and
        Instituto de Ciencias Exactas y Naturales (ICEN), Universidad Arturo Prat, Iquique, Chile
        \label{icen_chile}
        \and
        Université Paris-Saclay, CNRS, IJCLab, 91405 Orsay Cedex, France
        \label{ijclab}
        \and
        Caltech Optical Observatories, California Institute of Technology, Pasadena, CA 91125, USA
        \label{caltech_obs}
        \and
        The Oskar Klein Centre, Department of Astronomy, Stockholm University, Albanova University Center, Stockholm, SE-106 91, Sweden
        \label{okc_astro}
        }

  \date{}

  \abstract
  {The \textit{Zwicky} Transient Facility (ZTF) has been conducting a wide-field survey
    of the northern sky in three optical bands ($g$, $r$, and $i$). The ZTF
    collaboration is currently releasing light curves for 3628 spectroscopically
    confirmed Type Ia supernovae (SNe~Ia) discovered during its first 2.5 years
    of operation. This ZTF SN~Ia DR2 sample (DR2) is the largest SN Ia
    dataset to date, and provides an unprecedented opportunity to anchor the
    Hubble diagram.}
  {To fully exploit this dataset, our aim is to improve the accuracy of the
    light curve photometry to the 0.1\% level using a state-of-the-art scene
    modeling photometry (SMP) pipeline, which is optimal to extract a transient
    signal from a complex background (its host), while ensuring a common
    flux estimator with nearby stars used as calibration reference. However, the DR2 is
     based on forced photometry, which does not have such properties. For
    this paper, we compared our current SMP light curves with those released as
    part of the DR2 to assess the precision and accuracy of the ZTF SN~Ia DR2
    data.}
  {We have assembled a SMP pipeline that produces calibrated light
    curves for all DR2 supernovae (SNe). This pipeline is designed to
    efficiently process large datasets. The photometry is currently calibrated
    against the PS1 catalog, using SMP flux measurements of all the surrounding
    field stars in common as reference anchors.}
  {Our SMP pipeline can process the full ZTF 2.5-year dataset
    (\SI{179}{\tera\byte} of images) in about two weeks. In this paper we
    present the preliminary results obtained while producing an internal data
    release. Scene modeling light curves of the 3628 SNe~Ia in the DR2
    release were obtained in the $g$, $r$, and $i$ bands. The photometric
    repeatability of the observations is measured to be better than 1\%.
    However, in addition to a non-negligible brighter--fatter effect, we have
    identified a new sensor effect, dubbed  the pocket effect, which distorts the
    point spread function in a flux-dependent manner leading to
    nonlinearities in the photometry of up to 7\%. Correcting for this effect
    requires time- and sensor-dependent corrections to be applied at the pixel
    level. Currently, this issue prevents the existing data reduction pipelines
    from achieving the target photometric accuracy and affects all light curve
    releases to date,  from forced photometry (DR2) and from scene modeling
    (this work). We briefly discuss the origin of the effect and our plan to
    correct for it at the pixel level. Comparing the SMP and DR2 measurements,
    we find that stretch and color estimated from the two processings are
    consistent, aside from a \SI{10}{\milli\mag} shift in color. This assesses
    the robustness of results presented as part of the ZTF SN~Ia DR2 release.
    However, we find a global \SI{62}{\milli\mag} calibration offset between the
    two pipelines suggesting the absolute calibration of the DR2 pipeline is not
    known with a precision and accuracy needed to use these data for precise
    cosmological analyses. A reprocessing of the DR2 dataset using the SMP
    method is currently in progress.}
  {}


   \maketitle

\section{Introduction}
Type Ia supernovae (SNe~Ia) are among the most reliable and precise tools for
measuring cosmological distances. The detection of cosmic acceleration, which
led to the emergence of the standard model of cosmology, Lambda cold dark matter
($\Lambda$CDM), at the turn of the century, was initially accomplished using
fewer than 100~SNe~Ia at low and intermediate redshifts
\citep{perlmutterMeasurements42HighRedshift1999,
  riessObservationalEvidenceSupernovae1998,schmidtHighZSupernovaSearch1998}. In
the two decades that followed, SNe~Ia have played a central role in the
precision tests of $\Lambda$CDM, owing to the continuously expanding low- and
high-redshift SN samples suitable for cosmology
\citep{astierSupernovaLegacySurvey2006,kowalskiImprovedCosmologicalConstraints2008,rubinLookingLambdaUnion2009,
  sullivanSNLS3ConstraintsDark2011,suzukiHubbleSpaceTelescope2012,
  restCosmologicalConstraintsMeasurements2014,betouleImprovedCosmologicalConstraints2014,broutPantheonAnalysisCosmological2022}.
SNe~Ia are particularly crucial for investigating the recent phases of cosmic
acceleration, and consequently for constraining the equation of state (EoS) of
dark energy, $w = p / \rho$, along with its potential variations with redshift
($w_a$). Since the discovery of cosmic acceleration, a $\Lambda$CDM consensus
prevailed, and most analyses pointed toward values of $w$ compatible with $-1$
and $w_a$ of 0, \ie, with a cosmological constant.

Two state-of-the-art Hubble diagrams have recently been published, each
comprising approximately 2000 SNe in the redshift range $0.05 < z < 1$. The
first, Union3, is a compilation \citep{rubinUnionUNITYCosmology2025}, a large
composite dataset of 2087 SNe~Ia from 24 different surveys. The second was
published by the DES collaboration and includes the 5~year DES dataset at
intermediate and high redshifts (1635~SNe), complemented by a smaller sample of
194 low-redshift SNe from four different surveys
\citep{descollaborationDarkEnergySurvey2024a}. Intriguingly, both studies favor a
dark energy EoS in tension with $\Lambda$CDM, with a
higher-than-usual value of $w$ ($\sim -0.7$) and a nonzero drift with redshift.
However, we note that both measurements are not entirely independent since they
share a fraction of their low-redshift SN datasets. The recent DESI results
seem to indicate a similar tension: although the DESI dataset alone does not
show any deviation from $\Lambda$CDM, the combination of DESI and Planck also
favors a higher-than-usual value of $w$ \citep{adameDESI2024VI2025}. Jointly with
either Union3 or DES5yr, the tension between the $\Lambda$ values $w=-1$ and $w_a=0$ is at the
$\sim4\sigma$ level \citep{abdulkarimDESIDR2Results2025}. In that context,
building a SN Hubble diagram, entirely independent of the Union3 and DES
compilations seems crucial in order to assess the validity of this tension, largely
driven by SN Ia data, and especially by the low-redshift data points
\citep{abdulkarimDESIDR2Results2025}.

The \textit{Zwicky} Transient Facility (ZTF) SN Ia Data Release 2 (hereafter denoted ZTF
SN Ia DR2 or simply DR2) dataset \citep[][and references
  therein]{rigaultZTFSNIa2025a} comprises of 3628 Type Ia SNe in the redshift
range $0.02 < z < 0.09$. The dataset is spectroscopically complete up to $z \sim
0.06$ and has benefited from an exquisite photometric follow-up, using the ZTF
survey \citep{bellmZwickyTransientFacility2019} which aims to study the nearby
dynamic Universe \citep{grahamZwickyTransientFacility2019}. In terms of size,
completeness, homogeneity, and follow-up quality, it has no equivalent at low
redshifts. Combined with higher redshift samples it has the potential to anchor
the most constraining Hubble diagram before the statistics from the \textit{Rubin}
\citep{lsstsciencecollaborationLSSTScienceBook2009} and \textit{Roman}
\citep{spergelWideFieldInfrarRedSurvey2015} facilities become available.

As a rule of thumb, a deviation of $w$ of 0.01 from its standard $\Lambda$CDM
value of $-1$ results in a deviation of 0.2\% in the predicted distance modulus
with respect to its $\Lambda$CDM value in the most sensitive regions, namely
$z\sim 0.05$ and $z\sim 1$.  This gives an estimate of the precision required for
SN flux measurements. Consequently, it is essential to control every aspect of
the SN flux reconstruction, with a precision of the level of 0.1\%. This
includes (1) controlling the instrument's linearity, (2) controlling the
linearity of the SN flux estimator, (3) controlling the effective instrument
passbands, and (4) ensuring the accuracy of the flux metrology chain from the SNe
to the primary flux standards used to calibrate the surveys
\citep{bohlinTechniquesReviewAbsolute2014}.

We had two objectives when writing this paper. The first was to describe the
photometry methodology that is being deployed to analyze the ZTF dataset at this
level of precision. The second objective was to use the current results to
assert the precision and accuracy of the published ZTF SN Ia DR2 dataset.

The core of the method being developed for ZTF is an algorithm known as scene
modeling \citep{astierPhotometrySupernovaeImage2013,
  holtzmanSloanDigitalSky2008}, which consists of modeling the SN plus host
galaxy flux profile as the sum of a point spread function (PSF) trained on each
exposure and a galaxy model common to all exposures, convolved with an ad hoc
kernel to account for the seeing variations. This approach has two crucial
desirable properties. First, as a maximum-likelihood estimator, it is
statistically optimal \citep{astierPhotometrySupernovaeImage2013}. More
importantly, it can be applied ``as is'' to the field stars surrounding the SN.
This allows an external calibration to be explicitly propagated through the
field stars with a direct control, and at first order, auto-cancellation, of the
photometry related biases. In contrast, the ZTF SN Ia DR2 release is based on
forced photometry, \ie, on measuring the transient flux on difference images
made of flux-calibrated images. Hence, for this paper, we used the current
implementation of the scene modeling fluxes to assess the accuracy of the DR2
measurements. We also present our evaluation of the dataset's quality. We have
chosen not to release the scene modeling fluxes due to the discovery of a sensor
effect that induces flux-dependent variations in the PSF, leading to
nonlinearities in PSF fluxes (which also affect forced photometry fluxes). This
issue is in the process of being fixed, and scene modeling light curves will be
released at a later date.

In \Sect~\ref{sec:dataset} we describe the ZTF SN Ia DR2 dataset,
which was analyzed to produce the scene modeling light curves.
Section~\ref{sec:smp_pipeline} details the main aspects of our scene modeling
photometry (SMP) pipeline. In \Sect~\ref{sec:results} we present the main
results of our initial processing of the full DR2 dataset, including pipeline
performance, the first light curves, and an evaluation of basic quality metrics
particularly measurement repeatability based on field star light curves and
calibration precision. Section \ref{sec:comparison_forced_photometry} gives a
comparison between scene modeling and forced photometry light curves. In
\Sect~\ref{sec:limitations} we review the primary limitations that render these
light curves insufficient for cosmological analyses. A discussion on the sensor
effects discovered on the data impacting the linearity of the PSF is given,
along with an assessment of the impact of these effects on the photometric
measurements. We then evaluate the alignment of our band-pass models with stellar
measurements and discuss the requirements on band-pass models for cosmological
measurements. We conclude in \Sect~\ref{sec:conclusion}.

\section{Dataset}
\label{sec:dataset}

\subsection{The \textit{Zwicky} Transient Facility}
ZTF employs a \SI{47}{\square\deg} camera installed on the \SI{48}{\inch}
(\SI{1.2}{\meter}) Schmidt telescope at the Palomar Observatory to conduct a
high-cadence survey of the northern transient sky. This camera is equipped with
three filters, ZTF-$g$ ($g$), ZTF-$r$ ($r$), and ZTF-$i$ ($i$), and serves as a
dedicated instrument for the survey. With 30~s exposure times the ZTF observing
system is able to cover 3760~$\deg^2$ per hour with a typical depth of
\SI{20.5}{\mag} per visit in all three bands
\citep{bellmZwickyTransientFacility2019}.

A difference imaging pipeline is run at the end of every exposure and detections
are immediately transmitted into an alert stream
\citep{pattersonZwickyTransientFacility2019}. To systematically classify the
transients discovered with the imaging survey, ZTF uses a fully automated
low-resolution spectrograph, known as the Spectral Energy Distribution Machine
(SEDM; \citealt{blagorodnovaSEDMachineRobotic2018,
  rigaultFullyAutomatedIntegral2019, lezmyHyperGalHyperspectralScene2022}). It
reaches a spectroscopic transient classification completeness rate of 94\% at
$m_{\mathrm{peak}} < 18.5$ mag. Further away transients have had their spectra
secured through other telescopes \citep{perleyZwickyTransientFacility2020}.

\subsection{The ZTF telescope and camera}
The ZTF observing system is presented in detail in
\cite{dekanyZwickyTransientFacility2020}. The camera is a \SI{576}{\mega\pixel}
mosaic of 16 $6144\times 6160$ Teledyne/e2v CCD231-C6 CCDs. It covers the full
\SI{47}{\square\deg} field of view provided by the telescope design. The optical
design of the telescope had to be slightly modified, by adding an additional
corrector lens and one field flattener lens for each CCD, to compensate for the
disruptions caused by the cryostat window and the planar CCDs to the original
Schmidt design. The camera delivers a uniform image quality with a plate scale
of \SI{1.01}{\arcsecond} per \SI{15}{\micro\meter} pixel. Given the Palomar
median seeing of \SI{2}{\arcsecond}, the PSF is well sampled on roughly half the
exposures.

\paragraph{Sensors}
The sixteen sensors are organized in four rows of 4 sensors each. They have a
fairly large well capacity of about \SI{350000}{\electron}, and the gain of the
readout chain has been set at \SI{6.2}{\electron\per\ADU}, to compress the full
dynamical scale on the 16 bit ADU range. Each sensor is read out in
approximately \SI{9}{\second} through 4 different channels (64 readout channels
for the full camera), with a readout noise level near \SI{10}{\electron}.

The sensors of the upper and lower row (in declination) have a single layer
Anti-Reflective (AR) coating, while the sensors placed in the middle rows have a
dual layer AR coating, improving the response in the $g$ and $r$ bands. These
coating differences induce slight modifications in the quantum efficiency shape,
resulting in varying effective ZTF band-passes depending on the focal plane
position.

\paragraph{Band-passes} The three $g$-, $r$-, and $i$-broadband interference
filters are stored outside the telescope tube and mounted in front of the dewar
window using a robotic arm. The filter coatings have been characterized by the
vendor and found to be very uniform. However, the $f/2.4$ optical design induces
large variations in the incident beam angle as a function of focal plane
position, resulting in small continuous variations of the effective band-passes.
Combined with the variations in AR coating, this results in band-pass
non-uniformities of $\sim$\SI{3}{\nano\meter} (\resp $\sim$\SI{1}{\nano\meter})
peak-to-peak over the full focal plane for the $g$ and $r$ filters (\resp $i$).

\subsection{ZTF SN Ia DR2 dataset}
ZTF has been in operation since April 2018. The initial phases of the project
are referred to as ZTF-I covering the period from April 2018 to October 2020,
and ZTF-II, which extends from October 2020 to October 2023. The ZTF SN~Ia DR2
dataset presented in \cite{rigaultZTFSNIa2025a} (and hereafter noted DR2, not to
be confused with the ZTF DR2 data products unless explicitly specified) covers
the period from March 2018 to December 2020. It comprises of exclusively
spectroscopically identified SNe~Ia. The sample has been shown to be complete up
to a redshift of about $z=0.06$, with 994 objects \citep{amenoucheZTFSNIa2025}.

\subsection{SN Ia DR2 light curves}\label{sec:sn_ia_light_curves}

\paragraph{Forced photometry} The light curves of all the objects released in
DR2 have been obtained from the outputs of the ZTF pipeline
\citep{masciZwickyTransientFacility2019}. The DR2 dataset relies on a forced
photometry algorithm \CitepInPrep{Smith}, where the SN flux is evaluated by
fitting a PSF model to the difference images at the estimated location of the SN
; ZTF uses the ZOGY difference images algorithm presented in
\cite{zackayProperImageSubtraction2016}.

The DR2 light curves are used to perform an initial selection of the well
sampled SNe, using the SALT2 \citep{guySupernovaLegacySurvey2010} version T21
\citep{taylorRevisedSALT2Surface2021} light curve model, as exposed in
\texttt{sncosmo} \citep{barbarySNCosmo2024}. Out of the 3628 SNe in the DR2,
2960 are reported to have a good light curves by \cite{rigaultZTFSNIa2025a},
\ie, light curve with at least 7 detections in the $[-20,+40]$ day range
\citep{rigaultZTFSNIa2025b}, with 2 detections before peak and 2 after.

Approximately 15\% of the DR2 SNe are located in ZTF high-cadence
regions, typically yielding around 170 observations in the rest-frame phase
range of $[-10,+40]$ days in the $g$ and $r$ bands. The remaining 85\%, observed
at normal cadence, have approximately 30 observations over the same phase range.

Out of the 2960 well sampled SNe, 2667~SNe pass two additional cuts. The first
cut is designed to select SNe well sampled enough to accurately constrain the
SALT2 parameters: $(\sigma_{x_{1}} < 1)\ \&\ (\sigma_c <
0.1)\ \&\ (\sigma_{t_{peak}} < 1\ \mathrm{day})$. The second cut is designed to
select the bulk of the normal SN~Ia population: $(-3 < x_{1} < 3)\ \ \&\ \ (-0.2
< c < 0.8)$.

\paragraph{Limitations of forced photometry} The DR2 forced photometry light
curves are accurate at the 5\% level \citep{rigaultZTFSNIa2025a}. This is
perfectly adequate for SN population studies which is the main focus of DR2. All
the studies presented in the DR2 papers are based on these light curves
\citep[see][and references therein]{rigaultZTFSNIa2025a}.

However, for photometric flux estimates accurate at the per-mil level, forced
photometry suffers from three limitations, which degrade the accuracy of the
flux estimates:
\begin{enumerate}
  \item In current implementation of the ZTF image difference pipeline, the PSF
    used to estimate the flux on difference images is fixed and not spatially
    variable. Hence, it does not fully reflect the true PSF at the location of
    the SN.

  \item The position of the SN is pre-determined as the average of the positions
    reported by the detection pipeline as issued by the alerts
    \citep{bellmZwickyTransientFacility2019}. The algorithm then uses this
    average position to assert the RA, Dec coordinate of the SN. It then uses
    the WCS solution available from each ZTF image to force the PSF positions
    during the fit, solely leaving the PSF amplitude, \ie, the flux, as free
    parameter. This avoids biasing the flux estimation by position error in
    limited signal-to-noise images.
 
  \item The algorithm used to measure the SN cannot be applied as-is to the
    field stars as they are not present in the difference image. Yet, these
    stars are used to assert the photometric calibration. A direct anchoring of
    the SN force-photometry light curves calibration cannot thus be made on
    stars, but rely on assuming a cross-calibration between science images (with
    stars) and difference images (with transients). Yet, being able to apply 
    exactly the same estimator to the SN and field stars is a requirement for
    building an accurate flux calibration chain.
\end{enumerate}

These limitations have motivated the development and deployment of an alternate
photometry pipeline based on a scene modeling flux estimator that does not
suffer from the limitations listed above. It is more costly in terms of
computing time but specifically tailored for SN~Ia cosmology studies.
Nevertheless, it can be applied to a selection of the ZTF transients if needed.
In the remainder of this paper, we describe this effort aimed at providing
calibrated scene modeling light curves for the ZTF SN~Ia DR2 sample.

\section{Scene modeling photometry pipeline}
\label{sec:smp_pipeline}

We now describe the SN photometry pipeline constructed to deliver the scene
modeling light curves of the DR2 sample.

\subsection{The Scene modeling method}
\label{subsec:smp_pipeline:smp_method}

The scene modeling method we use is described in
\cite{astierPhotometrySupernovaeImage2013}. It consists in fitting a model of
the scene, \ie,~the SN and its host galaxy, on a series of image cutouts centered
on the SN. These include \textsc{on} images (containing SN light) and \textsc{off}
images (taken either before or after the SN has faded). The model is of the
form
\begin{equation}
  Y_{ip} = R_i \ \left[f_{i} \times \psi_{i}\left(\vec{x}_{p} - \vec{T}_{i}(\vec{x}_{\mathrm{SN}})\right) + K_{i} \otimes G\left(\vec{T}^{-1}_{i}(\vec{x}_{p})\right) + S_{i} \right],
  \label{eqn:scene_model}
\end{equation}
where the indices $i,p$ are running over the cutouts and cutout pixels,
respectively; $f_{i}$ is the SN flux on cutout $i$; and $\psi_{i}$ is the PSF model
on exposure $i$ at the position of the SN. The galaxy background flux profile,
noted $G(\vec{x})$ is modeled using a free pixel grid defined in the frame of
the best seeing cutout, called hereafter the reference cutout. To predict the galaxy
profile in cutout $i$, $G$ is convolved with $K_{i}$, a position-variable kernel
that allows   the reference and current PSF models to be matched: $\psi_{i} = K_{i}
\otimes \psi_{\mathrm{ref}}$. $\vec{T}_{i}$ are the coordinate mappings from the
reference cutout to image $i$, modeled as polynomial functions of the
coordinates. $S_{i}$ is an additive term introduced to account for sky level
differences between image $i$ and the reference image. Finally, the $R_{i}$'s
are the photometric coefficients that account for the throughput differences
between exposure $i$ and the reference exposure. We should note that the data is
not resampled, avoiding the introduction of pixel correlations.

Some model ingredients rely on information which is not present on the cutout
and are therefore determined before the scene modeling fit. This is the case of
the PSF models $\psi_{i}$, the convolution kernels $K_{i}$ which are directly
derived from the PSF, the relative geometric transformations $\vec{T}_{i}$ and
the photometric coefficients $R_{i}$. All are determined using the field stars
present on the full quadrant frames.

The model free parameters are (1) the SN fluxes $f_{i}$ (2) the SN position on
the reference $\vec{x}_{\mathrm{SN}}$ (whose initial estimates is given by
TNS-WIS and transformed through the reference quadrant WCS) (3) the galaxy model
$G$ and (4) the sky levels $S_{i}$, the only parameters of interest being the SN
fluxes. We note that the $S_{i}$ parameters are globally degenerate with the
galaxy model $G$, because adding a uniform flux to the galaxy model can be
offset by shifting all the $S_{i}$ parameters. For this reason, we fit for all
the $S_{i}$ parameters except one. The typical size of the parameter vector is
$\sim$1000, with typically $\sim$100 SN fluxes.

\subsection{Input data frames}

\begin{table*}
  \caption{Median number of exposures per SNe per cadence type along
      with the total number of frames to process compared to the full exposure
      dataset covered by the DR2.}
  \label{tab:input_frames}
  \begin{center}
    \begin{tabular}{lccccr}
      \midrule
      \midrule
      Band & High cadence     & Normal cadence   & Total SN      & Total DR2 & \% of DR2 \\
           & (per light curve) & (per light curve) &  (quadrants)  & (exposures/quadrants)          & \\
      \midrule
      $g$  & 853              & 217              &  \num{1.3e6}  &   \num{1.65e5} / \num{1.06e7}  & 12.3 \\
      $r$  & 901              & 231              &  \num{1.6e6}  &   \num{2.47e5} / \num{1.58e7}  &  9.5 \\
      $i$  &                  & 123              &  \num{3e5}    &   \num{1.9e4}  / \num{1.2e6}   & 25.0 \\
      \midrule
      \textbf{Total} &        &                  &  \num{3.2e6}  &    \num{4.31e5} / \num{2.76e7} & 11.6 \\
      \midrule
    \end{tabular}
  \end{center}
\end{table*}

We start our re-analysis of the ZTF SN~Ia DR2 dataset from the \texttt{sci}
exposures already detrended by the ZTF pipeline
\citep{masciZwickyTransientFacility2019}. The ZTF DAQ delivers 64 fits frames
per exposures, one for each amplifier. These frames are compressed (slightly
lossy Rice compression with \texttt{fpack}) at the telescope to accommodate the
limited bandwidth of the \textsc{hpwren} microwave network connecting Palomar
with the outer world. At IPAC, the resulting image is overscan corrected, bias
subtracted and flatfielded. Bias frames are taken daily and stacked using a
robust averaging algorithm to produce a master bias frame. The same process is
applied to create master flatfield frames from individual frames taken daily
with a LED-based illumination system. Bad pixels are identified and tagged
during this process.

Our effort uses (1) the detrended images for each quadrant (2) the mask images
that notably identify cosmics and saturated pixels and (3) the WCS solution
stored in the headers.

As detailed in Table~\ref{tab:input_frames} the volume of the ZTF exposure
dataset covered by the DR2 is substantial. It comprises 165k, 247k and 19k
exposures in the $g$, $r$, and $i$ bands, totaling 27.6~M individual quadrants.
Each quadrant \texttt{FITS} frame, is stored in floating point and occupies
about \SI{37}{\mega\byte}, along with a mask image of around
\SI{19}{\mega\byte}. Therefore, storing all the input pixels would require
approximately \SI{1.5}{\peta\byte} of disk space.

\begin{figure*}[t]
  \begin{center}
    \includegraphics[width=\linewidth]{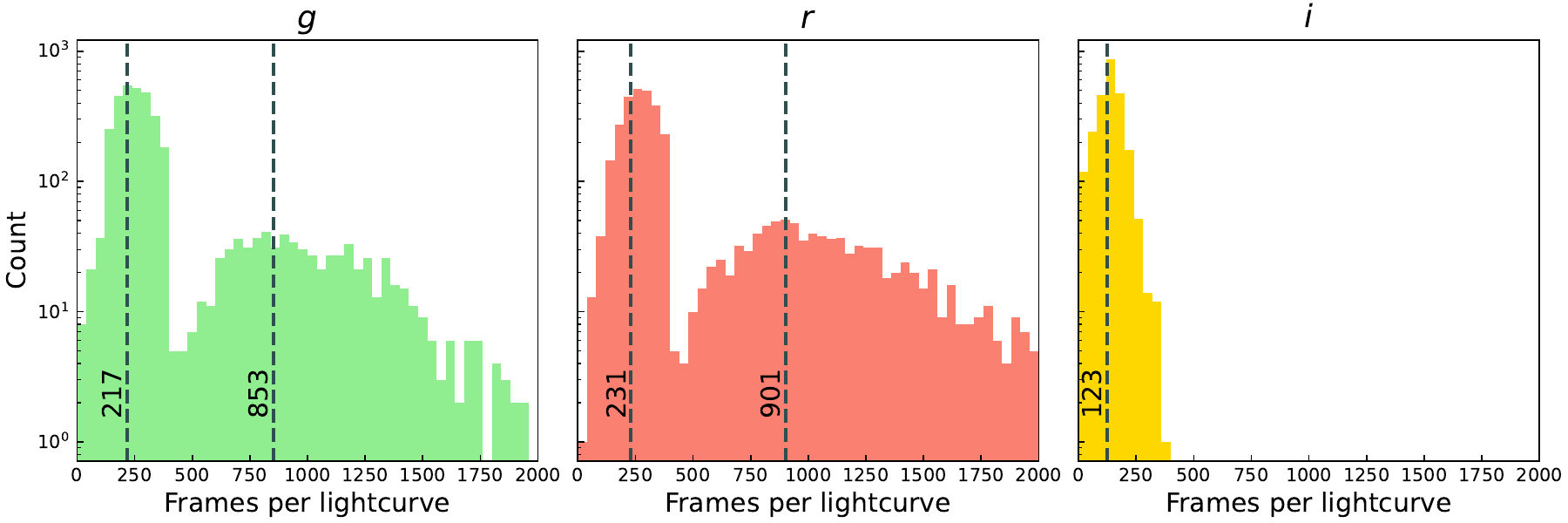}
    \caption{Total number of frames (\textsc{on} + \textsc{off}) per light curve
      in the $g$, $r$, and $i$ bands. The left (\resp~right) dotted vertical line
      represents the median exposure count per SN   in the normal-cadence
      (\resp~high-cadence) fields.}
    \label{fig:number_of_frames}
  \end{center}
\end{figure*}

In practice, we only need to process a small fraction of this large dataset. For
each SN, in each band, we first select all the quadrants that contain SN light,
called \textsc{on} frames. We complement these quadrants with at least 6 times
more (and at least 100) \textsc{off} quadrants, to help model the host galaxy
flux at the location of the SN. The \textsc{off} quadrants are selected from
exposures taken both before the SN explosion and once it faded. Specifically,
the \textsc{on} dataset contains all the quadrants containing the SN and taken
in the phase range $-50 < \Phi < +100$ (rest frame) days, while the \textsc{off}
data are selected from the exposures taken outside the phase range $-50 < \Phi <
+200$ days.

Figure~\ref{fig:number_of_frames} shows the number of (\textsc{on}+\textsc{off})
frames used in the scene modeling light curve processing in the $g, r$ and $i$
bands. In all bands except $i$, two distinct modes are visible, corresponding to
the normal- and high-cadence SNe. For a SN in the high-cadence fields,
approximately 500 to 1500 quadrant-frames need to be processed in $g$ and $r$
and about 120 in $i$. For a SN in the normal-cadence field, about 250
(\resp~120) frames enter the processing in $g$ and $r$ (\resp~$i$).
Table~\ref{tab:input_frames} summarizes these numbers. In total, for the full
DR2 dataset, we need to process about 3.2 million quadrants, \ie,~11.6\% of the
ZTF dataset. This still represents a sizeable volume of data:
\SI{179}{\tera\byte}.

\subsection{From pixels to catalogs}
\label{subsec:smp_pipeline:smp_pipeline_details}
We now describe the initial pipeline segments, run prior to the scene modeling
fit. First, the detrended quadrants delivered by the ZTF pipeline are
individually processed using \texttt{SExtractor}
\citep{bertinSExtractorSoftwareSource1996} to build an initial object catalog.
We use the segmentation frames also produced by \texttt{SExtractor} to produce a
refined background model, as immune as possible from object light. This model is
then subtracted from the image.

A map of pixel weights is then constructed. It contains the inverse sky
variance, modulated by the flatfield frames. It does not include the
contribution of the object fluxes to the variance, as this would bias the point
source PSF flux estimates \citep{astierPhotometrySupernovaeImage2013}.

The initial \texttt{SExtractor} catalog is matched with the intersection of the
\textit{Gaia}-eDR3
\citep[\textit{Gaia};][]{gaiacollaborationGaiaMission2016,gaiacollaborationGaiaEarlyData2021}
and Pan-STARRS \citep[PS1;][]{magnierPanSTARRSPhotometricAstrometric2020}
catalogs of the ZTF field. The objects identified as stable stars in the \textit{Gaia}
and PS1 catalogs, using a combination of PS1 flags are retained as input to the
PSF training process.

To save processing time, we do not re-compute at this stage the astrometric
solutions determined by the ZTF/IPAC pipeline and stored in WCS format in the
image headers. Their precision, of the order of 50~milli-arcsecond (mas), is
largely sufficient to associate sources from external catalogs. However, since
astrometric uncertainties transform into photometry biases, we do refine the
relative image-to-image transformations ($\vec{T}_{i}$ in
\Eq~\ref{eqn:scene_model}) prior to each SMP fit, using the IPAC WCS transforms
as a starting point (\Sect~\ref{subsec:smp_pipeline:astrometric_transforms}).

At this stage, the detrended, sky-subtracted pixels are ready to be injected
into the scene modeling code. The best-seeing image among the ZTF-primary grid
exposures is selected as a reference image. Cutouts centered on the SN,
encompassing the SN+Galaxy scene are extracted from the stack of \textsc{on}
and \textsc{off} exposures. The size of each cutout for exposure $i$ is
determined by extending its PSF, $\psi_{i}$, with the half-size of the
associated kernel, $K_{i}$ (see \Sect~\ref{subsec:smp_pipeline:psf_model}
and~\ref{subsec:smp_pipeline:kernels} for details on the PSF and kernel
determinations). The galaxy model, sampled at the same resolution as the
exposure, has its size defined by the union of all cutout coverages, transformed
via $T_{i}$ into the reference exposure's frame. For a typical light curve, the
model spans a $30 \times 30$ pixel area.

We now detail the determination of the key model ingredients which are (1) the
PSF models ($\psi_{i}$, \Sect~\ref{subsec:smp_pipeline:psf_model}), the
frame-to-frame astrometric transformations ($\vec{T}_{i}$,
\Sect~\ref{subsec:smp_pipeline:astrometric_transforms}) and the photometric
ratios ($R_{i}$, \Sect~\ref{subsec:smp_pipeline:photometric_alignment}). These
quantities rely on the stars identified on the full frames, before the cutouts
are extracted.

\subsection{PSF model}
\label{subsec:smp_pipeline:psf_model}
The PSF modeling code was originally written specifically for the Supernova
Legacy Survey (SNLS), and also used on the Subaru/HSC high-redshift datasets.
The technique is inspired by \texttt{DAOPhot}
\citep{stetsonDAOPHOTComputerProgram1987} and documented
in~\cite{astierPhotometrySupernovaeImage2013}. The trained model is the sum of a
(position-variable) analytical core (generally a Moffat) and a pixel-grid, to
capture the remainder of the PSF profile. The pixel grid part of the model is
also variable as a function of position. On each typical exposure, 400 to 1000
stars, in the magnitude range $14 < g < 20$, can be used to train the PSF model.

As discussed in~\cite{astierPhotometrySupernovaeImage2013}, the PSF model is
trained with pixel weights accounting for all sources of noise, including the
object flux itself, following
\begin{equation*}
w_{i}^{-1} = \mathrm{Var}\left[f_{\mathrm{sky}}\right] + \frac{1}{g} f \psi_i,
\end{equation*}
where $f_{\mathrm{sky}}$ is the sky background flux, $g$ the amplifier gain, $f$
the star flux, and $\psi_i$ the value of the model at position $i$. This has the
disadvantage to make the relative weights of the pixels entering the fit
dependent on the PSF model itself, and the ratios of flux estimates of bright
and faint source also dependent of the PSF model. To mitigate this problem, we
redetermine the star fluxes after training, retaining only the sky variance
component for the weights: $w_{i}^{-1} = \mathrm{Var(sky)}$. This ensures that
we have  exactly the same estimator for faint and bright sources including the SN
itself, while remaining statistically optimal for the faint SNe. In order to
minimize the adverse effects of undersampled images (which is the case with a
large part of ZTF exposures), the analytical part of the PSF is integrated over
the area of each pixel using a 3-point Gaussian quadrature rule.

\subsection{Astrometric transforms}\label{subsec:smp_pipeline:astrometric_transforms}
We determine the astrometric transforms $\vec{T}_{i}$ from the reference cutout
to each cutout of the stack as follows.

We use the \textit{Gaia} catalog as an astrometric reference. The positions and proper
motions of the \textit{Gaia} stars are projected on the tangent plane centered on the
position of the SN. Polynomial geometric transformations $\vec{P}_{i}$ are then
computed to map coordinates from this tangent plane (in angular units) to the
pixel coordinates of each quadrant frame, accounting for the proper motion of
each field star. The $\vec{T}_{i}$ mapping are then determined as
\begin{equation}
  \vec{T}_{i} = \vec{P}_{i} \circ \vec{P}^{-1}_{\mathrm{ref}}.
\end{equation}
The inversion and composition of the $\vec{P}_{i}$ transforms are performed
numerically. We generate an adapted grid $g^{\mathrm{TP}}_{pq}$ on the tangent
plane. This grid is transported into each quadrant frame using the $\vec{P}_{i}$
transformations: $g^{i}_{pq} = \vec{P}_{i}\left(g^{\mathrm{TP}}_{pq}\right)$.
The coefficients of the $\vec{T}_{i}$ transforms are fit directly on the
$\left(g^{\mathrm{ref}}_{pq}, g^{i}_{pq}\right)$ grids. We have verified that
provided the grid is dense enough, these inversion and composition operations
are precise at a fraction of a tenth of a mas on the full frame.

\begin{figure}[t]
  \begin{center}
    \includegraphics[width=\linewidth]{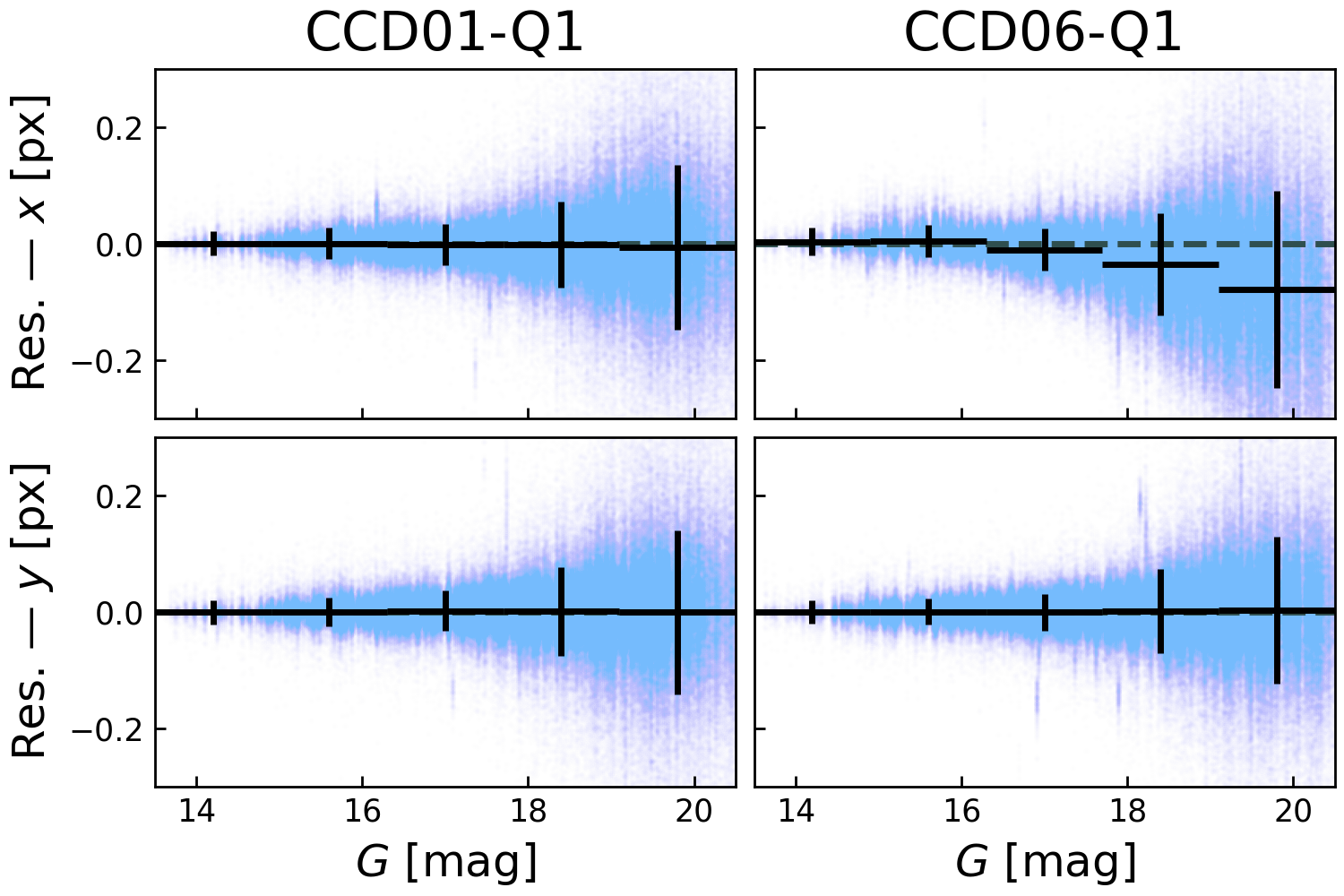}
    \caption{Astrometric residuals, in $r$ band, on quadrant 1 of CCD 1 (\textit{left})
      and CCD 6 (\textit{right}), in the $x$- (\textit{top}) and $y$-direction (\textit{bottom}), as a function of \textit{Gaia} magnitude $G$. The horizontal gray
      dotted lines show zero. For bright stars, the residual dispersion is about
      \SI{0.08}{\pixel} (\SI{80}{\milli\arcsec}) and \SI{0.04}{\pixel}
      (\SI{40}{\milli\arcsec}) in the $x$- and $y$-directions respectively. We
      note that for some sensors (here CCD 6) the $x$-direction residuals
      display a trend as a function of flux. This is one of the manifestations
      of the pocket effect (see \Sect \ref{subsec:limitations:sensor_effects}).}
    \label{fig:astrometric_residuals}
  \end{center}
\end{figure}

We show in \Fig~\ref{fig:astrometric_residuals} typical astrometric residuals
obtained by the chain, for two different CCDs. In the top right panel of the
figure, the residuals in the $x$-direction (which corresponds to the serial
direction of the sensor) strongly depend on the star flux. This is one of the
manifestations of the new, CCD dependent, sensor effect dubbed the pocket
effect, identified in our first pass on the ZTF dataset, which distorts the
PSF shape in the serial direction as a function of the star flux. We discuss
this effect and the mitigation strategies that were developed in
\Sect~\ref{subsec:limitations:sensor_effects}.

We observe from \Fig~\ref{fig:astrometric_residuals} that the typical
astrometric precision obtained on the quadrant frames for bright stars is of
about \SI{0.08}{\pixel} (\SI{80}{\milli\arcsec}) in the $x$-direction, and
\SI{0.04}{\pixel} (\SI{40}{\milli\arcsec}) in the $y$-direction. This gives an
estimate of the precision of the astrometric transformations used in the scene
modeling. Once again, the difference in precision between the $x$- and
$y$-directions is attributed to the pocket effect.

Astrometric precision is an important quantity as astrometric uncertainties
translate into a bias on the estimated fluxes. For a Gaussian PSF,
the uncertainty $\Delta f$ follows
\begin{equation}
\frac{\Delta f}{f}=\frac{1}{4}\frac{\delta x^2+\delta y^2}{\sigma_{\mathrm{seeing}}^2},
\end{equation}
with $\delta x$, $\delta y$ respectively the $x$- and $y$-direction
uncertainties and $\sigma_{\mathrm{seeing}}$ the image seeing. With a typical
astrometric precision of 80 and 40 mas in the $x$- and $y$-directions, however,
and a typical seeing of \SI{2.3}{\arcsecond} FWHM, we estimate that the
astrometry-related bias affecting the SN fluxes is of the order of
2~milli-magnitudes (mmag). It reaches \SIrange{3}{4}{\milli\mag} on good seeing
ZTF exposures ($\sim$\SI{1.5}{\arcsecond}). The average value of this bias is
entirely absorbed by photometric calibration. We note however that the increase
in astrometric dispersion as a function of magnitude translate into a
magnitude-dependent bias, estimated about \SI{2}{\milli\mag} peak-to-peak over
the \SIrange{14}{20}{\mag} range.

\subsection{Photometric alignment}\label{subsec:smp_pipeline:photometric_alignment}
The photometric scales, $R_{i}$ correct for nonchromatic variations of the
telescope throughput between the reference image and each exposure: $f_{ij} =
R_{i} \times f_{i_{\mathrm{ref}}j}$, where $f_{ij}$ (\resp $f_{i_{ref}j}$) is
the PSF flux of star $j$ on exposure $i$ (\resp on the reference exposure).

The photometric scales are determined prior to the scene modeling fit using the
PSF fluxes of the field stars. Nearly all ZTF SNe are observed in more than one
quadrant, due to the overlapping of the ZTF fields. As a consequence, the
intersection between the reference quadrant and some of the other quadrants may
be low, with only a few objects in common (in some cases fewer than ten). For
this reason, the photometric scales are determined using a global fit, combining
all the frames entering the scene modeling fit. Since the fit becomes linear
when expressed in magnitude, we convert the PSF flux measurements into
instrumental magnitudes $m = -2.5 \log_{10} f$ and we fit the following model
simultaneously on all the exposures,
\begin{equation*}
  m_{ij} = m_{j} + zp_{i},
\end{equation*}
where the $j$ index runs over the stars and the $i$ index runs over the
exposures. The reference frame zero-point $zp_{i_{ref}}$ is fixed to zero. We
implement a robust fitting procedure, where outliers are identified and removed
in incremental steps. After the fit, the $m_{j}$ are marginalized over, and the
photometric scales are determined from the $zp_{i}$ estimates:
$R_{i} = 10^{0.4 (zp_{i} - zp_{i_{\mathrm{ref}}})}$.

\begin{figure}
  \begin{center}
    \includegraphics[width=\linewidth]{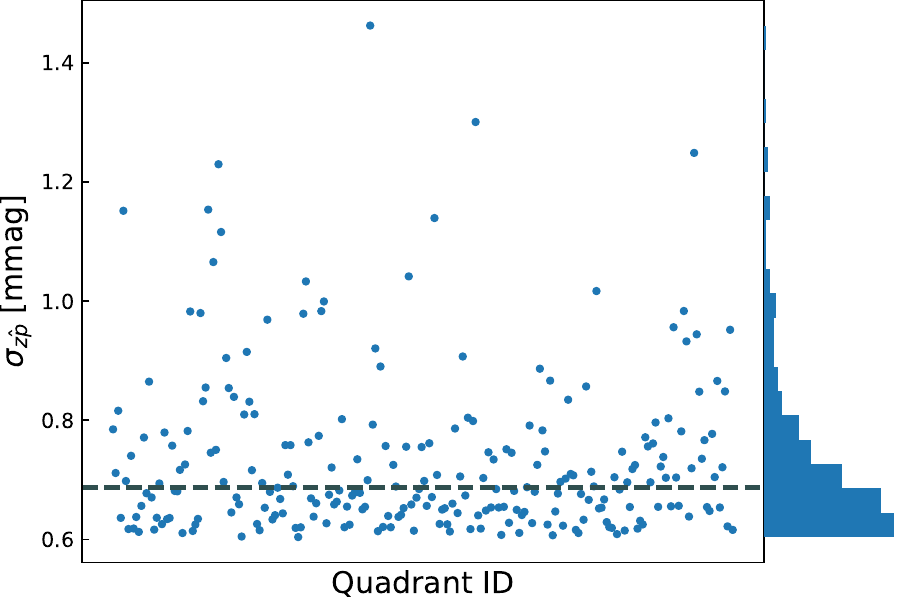}
    \caption{Uncertainties on the relative calibration offsets
      ($zp_{i}-zp_{i_{\mathrm{ref}}}$) in $g$ band CCD~6. The horizontal gray
      dotted line shows the median value.}
    \label{fig:photo_dzp}
  \end{center}
\end{figure}

We show in \Fig~\ref{fig:photo_dzp} the typical uncertainties affecting the
fitted $zp_{i} - zp_{i_{ref}}$ parameters. As can be seen, the typical
uncertainties are less than \SI{0.8}{\milli\mag}.

\subsection{Kernels}\label{subsec:smp_pipeline:kernels}
Finally, we need to determine the convolution kernels $\vec{K}_i$ which connect
the PSF of the reference quadrant with the PSF of any exposure $i$ entering the
fit. These kernels are notably used in the model, to predict the host galaxy
background shape on cutout $i$ from its modeled shape on the reference cutout.
These kernels are position dependent, and follow
\begin{equation}
  \psi_{i} = K_{i} \otimes \psi_{\mathrm{ref}}.
\end{equation}
The kernels are modeled as a free pixel grid. Each pixel value is modeled using
a polynomial of the position. They are fitted directly on a grid of PSF model
instantiations, as described in~\cite{astierPhotometrySupernovaeImage2013}.

\subsection{SN and field star light curves}
\label{subsec:smp_pipeline:smp_lc}
\begin{figure}[h]
  \begin{center}
    \includegraphics[width=\linewidth]{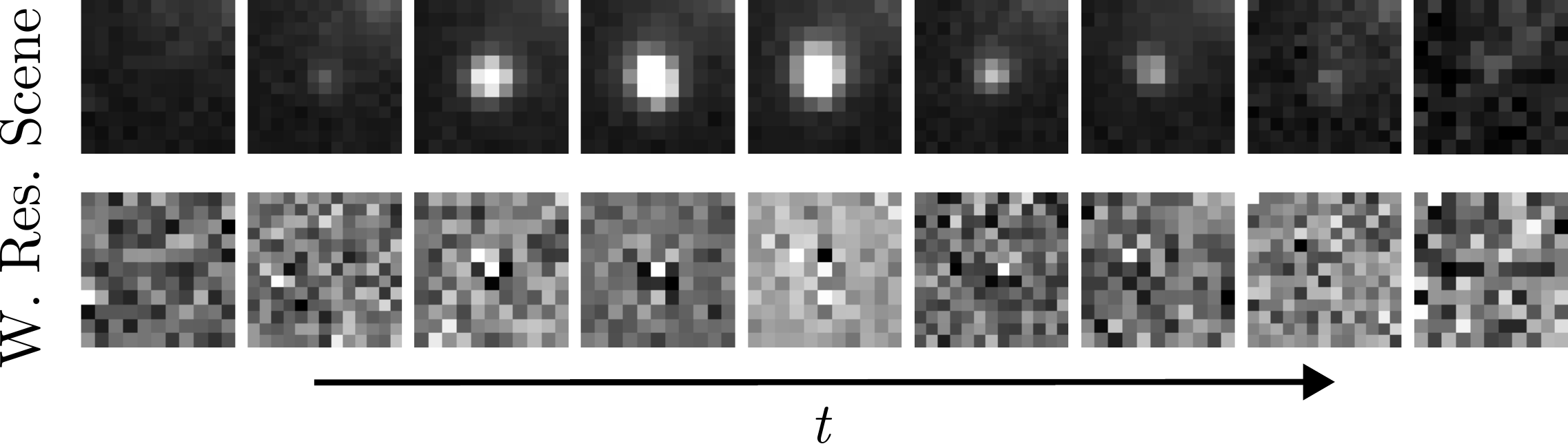}
    \caption{Vignettes of the SMP model (top) and weighted residuals (bottom)
      for SN2019ecs (ZTF19aaripqw) in $g$ band.}
      \label{fig:scene_model}
  \end{center}
\end{figure}

At this stage, we have all the necessary components of the model and can proceed
with the scene modeling fits. Since the bands are independent, a separate fit is
performed for each SN in each band. The fitting process employs a robust
minimization algorithm to handle the outlier pixels that are inevitably present
in the dataset. Upon completion of the minimization, the primary parameters of
interest --the scene modeling fluxes-- are extracted and stored, along with the
Fisher matrix uncertainties provided by the fitting procedure. Additionally, the
galaxy profile model is saved as a \texttt{FITS} file for later inspection, as
for the scene sequence (\Fig~\ref{fig:scene_model}). This model is also used to
estimate the host galaxy's local colors at the SN position. On an AMD EPYC 7302
CPU (3 GHz), the (single core) time required for each SN light curve fit
typically ranges from several tens to a few hundred seconds, depending on the
number of quadrants included in the fit, as shown in
\Fig~\ref{fig:pipeline_timings}.

We also generate scene modeling light curves for all stars present in the
quadrants involved in the fit. The star fits differ from the SN fits only in
that the galaxy background is fixed at zero. All other components (PSF,
astrometric transformations, flux scales, kernels, and scene model) are kept
identical. With the galaxy background fixed, the free parameters of the fit are
a global position and a flux value per exposure. These field star light curves
are utilized to evaluate the performance of the scene modeling fit, specifically
to assess the photometric repeatability and to verify that the uncertainties
reported by the fit accurately represent the observed repeatability for stable
stars. The averaged fluxes of field stars are subsequently used in the light
curve calibration chain (as described in
\Sect~\ref{subsec:smp_pipeline:smp_lightcurve_calibration}). The fitting process
for field stars requires between a few $10^{3}$ and a few $10^{4}$~s, depending
on the number of quadrants and the stellar density in the field, as illustrated
in \Fig~\ref{fig:pipeline_timings}.

\subsection{Light curve calibration}
\label{subsec:smp_pipeline:smp_lightcurve_calibration}

\paragraph{Flux metrology chain} The calibration of the SN light curves
ultimately relies on primary standards, which are spectrophotometric standards
with spectra accurately determined in physical units. Supernova-based cosmology
measurements, $w$, $H_{0}$ and $f\sigma_{8}$, are insensitive to absolute flux
scale of the survey. On the other hand, SN constraints on the dark energy
EoS are almost degenerate with the (relative) intercalibration of
the survey bands. Consequently, the critical calibration information is actually
encoded in the shape of the Primary standard spectra, which determines
the survey's band-to-band calibration.

Several sets of primary standards are available, with the state of the art
currently being the CALSPEC library \citep{bohlinNewGridsPurehydrogen2020}. The
CALSPEC flux scale is derived from models of a set of DA white dwarfs, whose
temperature and surface gravity are constrained from high-resolution
spectroscopic observations. While CALSPEC primary standards are approximately
\SI{4}{\mag} brighter than the typical ZTF SNe, most are within the ZTF
footprint and not saturated in \SI{30}{\second} exposures. A large effort is
underway to build a short metrology chain linking CALSPEC standard stars to SN
light curves relying exclusively on ZTF observations \CitepInPrep{Racine}. The
output of this effort will be a full sky catalog, with controlled uniformity,
and fluxes calibrated in units of the primary standard star observations. While
the ZTF calibration catalog is still in completion, we use the PS1 catalog as a
calibration reference for this first pass on data.

The calibration process relies on the field stars surrounding each SN. By
comparing the instrumental magnitudes of these field stars with their calibrated
catalog counterparts, a zero point is determined and subsequently applied to the
SN light curves. Since the zero-point is dependent of the flux estimator used,
it is crucial to apply the same flux estimator to the SNe and field stars. This
is the primary reason for using SMP instead of forced photometry on image
subtractions. In addition to SN light curves, the scene modeling pipeline
therefore generates light curves for all surrounding field stars with
counterparts in the PS1 catalog. Variable stars are identified and excluded from
the calibration process. The light curves of stable stars go through a robust
averager, producing an average scene modeling flux $\hat{f}_{\mathrm{SMP}}$.

\section{Results}
\label{sec:results}

\subsection{Pipeline implementation and performance}
\label{subsec:results:smp_implementation}

Extracting the scene modeling light curves from the pixels turned out to be a
challenging task. As discussed in \Sect~\ref{sec:dataset}, the information is
embedded within 3.2~M quadrant frames, representing about \SI{179}{\tera\byte}
of data. Although less daunting than the upcoming LSST processings, this is
still about two orders of magnitude larger than previous projects, and was
considered as major risk for this project. Additionally, a key requirement for
this pipeline was that it had to be able to process the full DR2 dataset in less
than a month, thus allowing   several iterations on the full dataset within a year.

\begin{figure*}[t]
  \begin{center}
    \includegraphics[width=\linewidth]{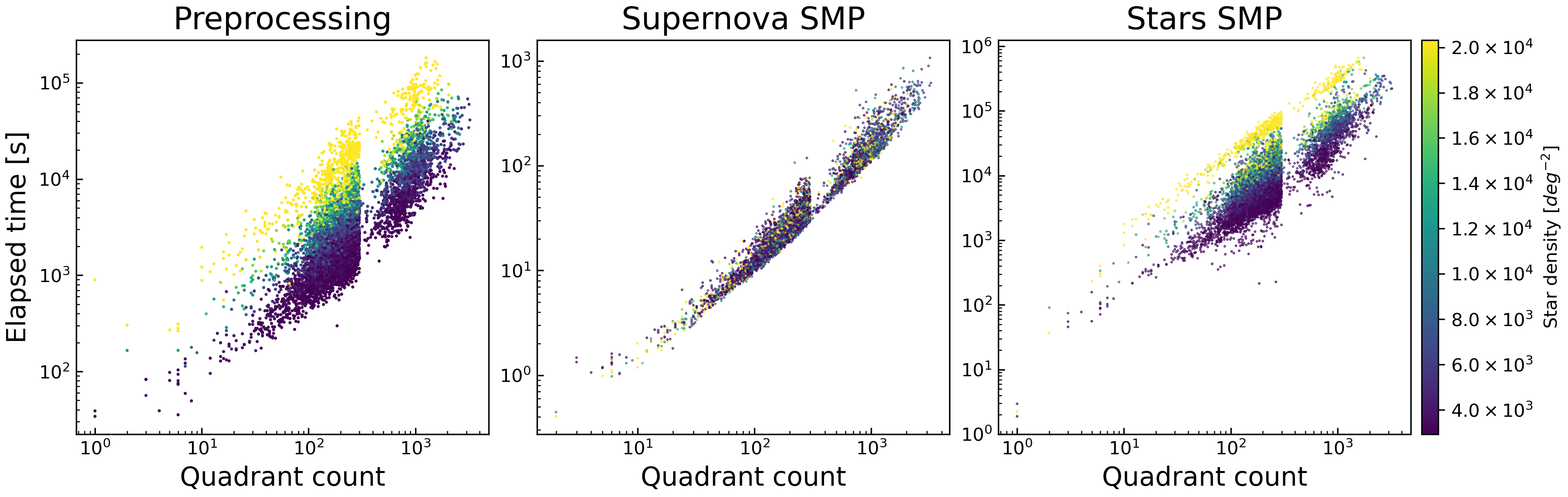}
    \caption{Typical durations of the processing steps, as a function of the
      number of quadrants entering the processing and as a function of the star
      density (color-coded).}
    \label{fig:pipeline_timings}
  \end{center}
\end{figure*}

The pipeline has been deployed at the \texttt{CC-IN2P3} computing center, which
allocated \SI{1.2}{\peta\byte} of disk space to the project and can provide up
to 2000~cores simultaneously for a full reprocessing of the dataset. The
computing farm is shared with other projects, notably LHC experiments and uses
\texttt{slurm} as a job scheduler.

Although \SI{1.2}{\peta\byte} is a large amount of disk space, it is not
sufficient to store, along with the input data, the entirety of the intermediate
products (\eg, reduced pixel frames, pixel weights, star catalogs, light
curves), which consume three to four times the input data volume.
Therefore, each job was designed to store the intermediate products only
temporarily on the local worker scratch disk space. By default, only the SN and
field star light curves are written to permanent disk storage. Intermediate
products are stored on demand, for a small number of objects, for debugging
purposes only. Since there are never more than one SN on a given quadrant frame,
this strategy does not introduce any inefficiencies. This will change in the
future, when we decide to process the entirety of the photometric dataset, which
is much denser spatially.

Each job processes one SN light curve in one single band independently, with
approximately 9,700 jobs scheduled to reprocess the entire DR2 dataset. A job
takes as an input all the quadrant frames needed to produce the light curve (see
Table~\ref{tab:input_frames}), processes them on the fly (detrending, object
detection, sky subtraction, PSF modeling, followed by the relative astrometric
transforms and the determination of the photometric scales), extract the
cutouts, runs the scene modeling fit, for the SN and the field stars, and dumps
the SN and field star light curves on permanent disk storage. Within the jobs,
all the tasks described above are scheduled internally using the
Dask\footnote{\url{https://www.dask.org}} framework
\citep{daskdevelopmentteamDaskLibraryDynamic2016}.

The core of the photometry pipeline, \ie,~the PSF model and the scene modeling
loop, reuses the C++ code developed for SNLS and is described
in~\cite{astierPhotometrySupernovaeImage2013}. All the other parts of the
pipeline (astrometry, photometric scales, calibration, job logical structure)
are written in Python.

Figure \ref{fig:pipeline_timings} presents the typical duration of the jobs and
their main subcomponents. As observed, there are two distinct classes of jobs:
those associated with the normal-cadence and high-cadence SNe respectively.
Unsurprisingly, the typical duration of the primary pipeline operations scales
linearly with the number of quadrants processed. The pre-processing steps
which encompass all the operations from input data frames to PSF modeling
require approximately \SI{10}{\second} per quadrant, for fields of normal
stellar density. This duration can increase by up to a factor of ten in regions
of high stellar density, such as those near the Galactic plane. The time
required for the SN scene modeling is minimal, ranging from 10 to 100~s for
normal- and high-cadence SNe, respectively. The most time-consuming step is the
scene modeling fit of field stars, which typically takes two to three times
longer than the preprocessing phase.

Figure \ref{fig:pipeline_timings} indicates that a normal-cadence object, with
approximately 100~frames, is processed in typically 1.5~hours (from pixels to
light curves, including field star light curves). In contrast, a high-cadence
object with 1000~frames requires about 11~hours of computing time. Depending on
the availability of the computing center, we are able to reprocess the full
dataset in two to three weeks, well within our initial requirements.

\subsection{Light curves}
\label{subsec:results:smp_lightcurves}
\begin{figure}
  \begin{center}
    \includegraphics[width=\linewidth]{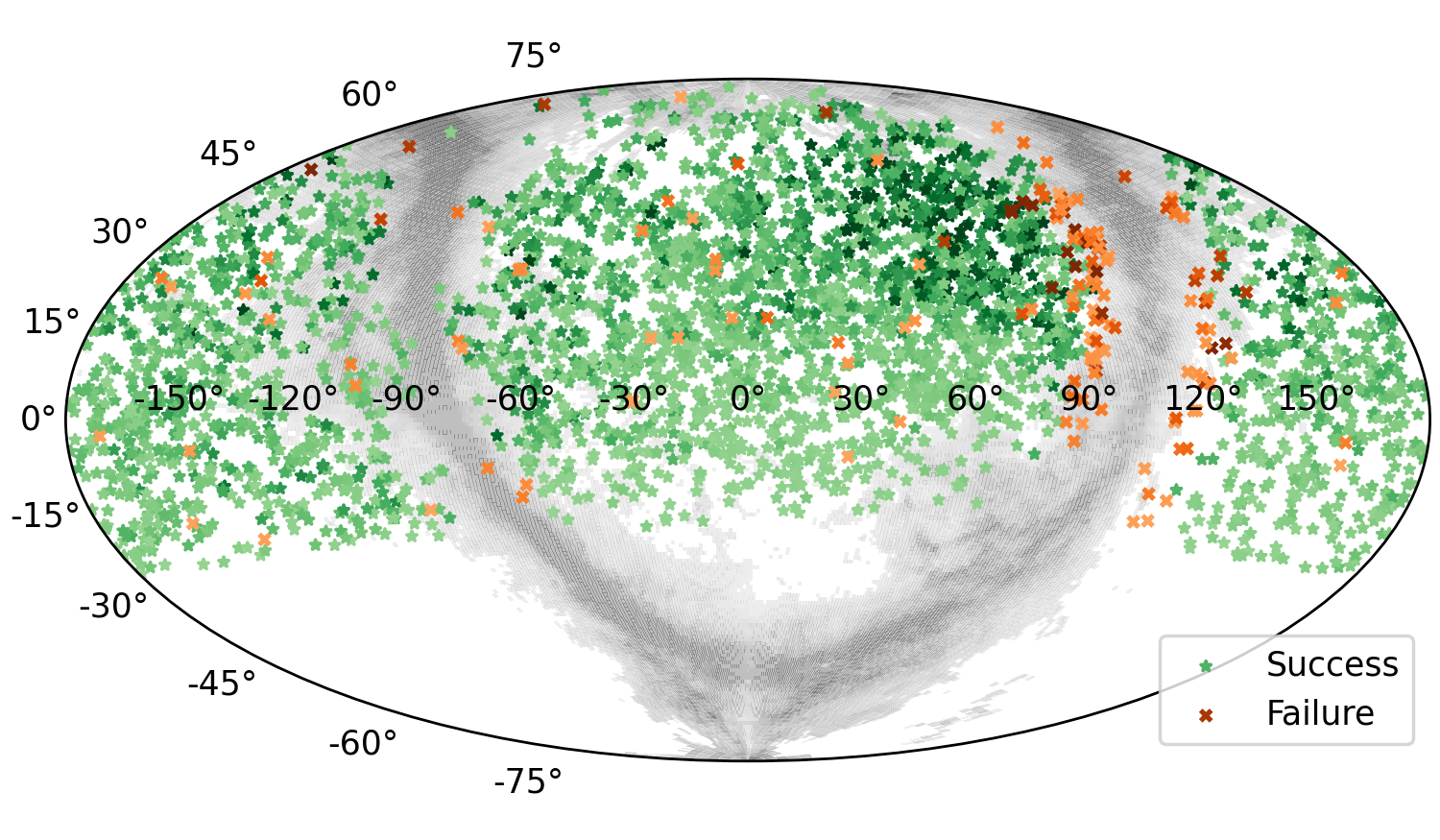}
    \caption{Distribution of the SMP SNe~Ia within the ZTF footprint. The SNe
      successfully processed by the pipeline are shown in green, while those for
      which the pipeline failed are depicted in red. Darker markers denote
      SNe with more points (> 1000) in their light curves. The
      majority of the failed cases are located near the Galactic plane, where
      the high stellar density poses significant challenges for the pipeline.
      These SNe are also likely to be affected by substantial Galactic
      extinction (shown in gray) \citep{schlaflyMeasuringReddeningSloan2011}, and
      will not be retained in the final dataset.}
    \label{fig:dr22_fails}
  \end{center}
\end{figure}

To date, two full iterations over the full dataset have been completed. The
first iteration took two weeks, while the second took three weeks to finish. The
difference in processing time was primarily due to varying workloads at the
CC-IN2P3 computing center. Each iteration resulted in an internal data-release.

Each processing includes the light curves of 3582~SNe (out of 3826), from which
9493 (out of 9756) light curves were produced. Of these, 1623 are classified as
high-cadence SNe, with typically 170 observations in the restframe phase
range of $[-10,+40]$ days. The remaining 7870 are normal-cadence objects, with
about 30 observations in the same phase range. About 2.6\% of the light curves
could not be processed, primarily due to high stellar density in the field,
which exceeded the pipeline's processing capabilities. Figure \ref{fig:dr22_fails}
displays the distribution of the SNe in the ZTF footprint, with failed cases
marked in red. We note that these failures are predominantly located near the
Galactic plane, where they are also likely affected by significant Galactic
extinction. As a result, they have been excluded from the final cosmological
dataset. The second most common cause of failure is a very low light curve
sampling, exclusively in the $i$ band, for which SMP cannot converge. As their
forced photometry-based light curves have not been produced (and thus not part
of DR2); those are discarded.

\subsection{Repeatability}\label{subsec:results:repeatability}
\begin{figure}[h]
  \begin{center}
    \includegraphics[width=\linewidth]{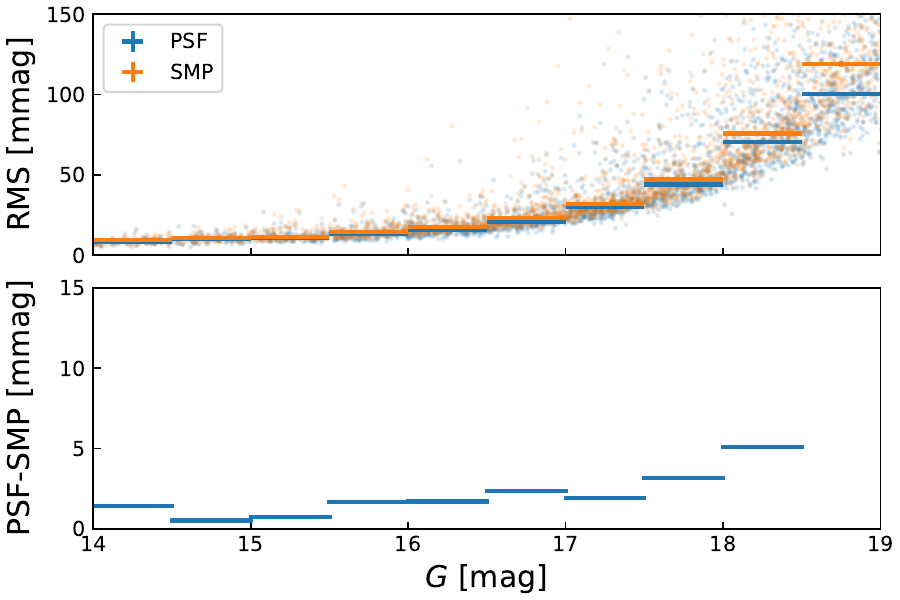}
    \caption{Field stars light curve RMS in $g$ band. The top plot shows the RMS
      value of each star light curve, computed through either free PSF or SMP,
      along with their binned values. The bottom plot shows the difference of
      each binned value, indicating a floor \SI{2}{\milli\mag} precision deficit
      for SMP in respect to PSF. Most SNe lie toward the right-hand side of the
      plot, at fainter magnitude, and are thus more sensitive to
      astrometry-related photometry bias.}
    \label{fig:repeatability}
  \end{center}
\end{figure}

The image-to-image repeatability of the SMP photometry can be directly assessed
as the (robust) root mean square (RMS) of the SMP light curves of field stars.
This is a crucial metric, as it allows us to evaluate the uncertainty floor, and
validate the photometric uncertainties reported by the pipeline. Figure
\ref{fig:repeatability} presents the weighted root mean square (wRMS) of the
stars' SMP light curves compared to free PSF light curves --where the star
position is fitted independently on each exposure
(\Sect~\ref{subsec:smp_pipeline:photometric_alignment}). Variable stars are
excluded and the wRMS is shown as a function of star magnitude. We observe that
(1) the repeatability of the free-PSF photometry is of the order of
\SI{10}{\milli\mag} for the brightest stars and (2) the SMP RMS is slightly
higher, by approximately \SI{2}{\milli\mag}. This higher RMS comes from the fact
that in the SMP model, the position is fixed to the prediction of the
astrometric transforms. It reflects the impact of astrometric noise as discussed
in \Sect~\ref{subsec:smp_pipeline:astrometric_transforms}.

Since a given quadrant possess in average over 700 independently measured stars,
its zero-point estimation, following a $1/\sqrt{N}$ law, is estimated to be less
than 1 mmag (as shown is \Sect~\ref{subsec:smp_pipeline:photometric_alignment}).
As a result, with the addition that SNe in the ZTF DR2 are well sampled in time,
contribution of the photometric error floor to their distance modulus become
subdominant to their intrinsic variability.

\subsection{Calibration}\label{subsec:results:smp_calibration_results}

\begin{figure*}[t]
  \begin{center}
    \includegraphics[width=0.9\linewidth]{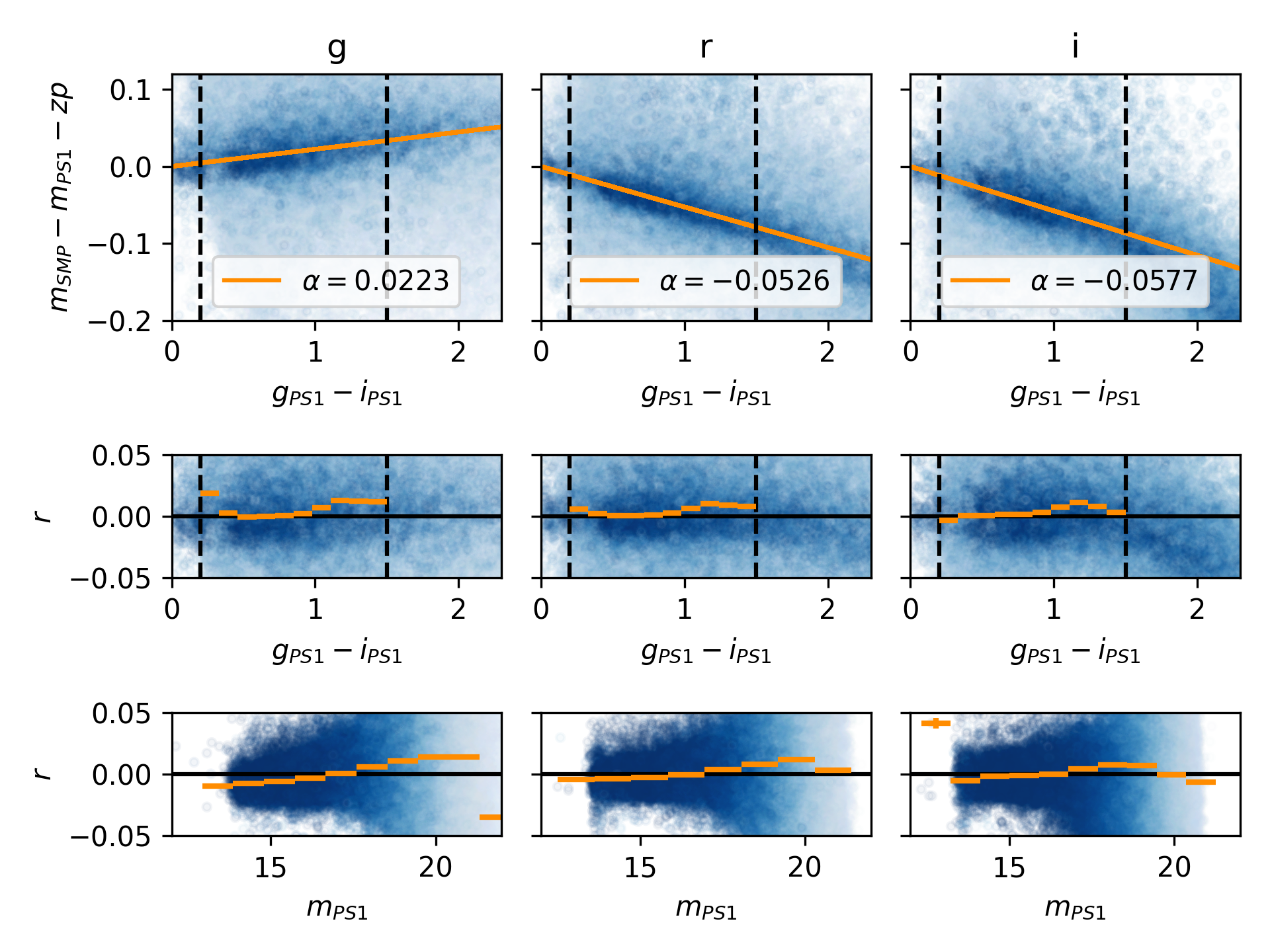}
    \caption{Top panel: Color transformation between ZTF and PS1 fitted on all
      stars (orange line). Raw measurements are shown as blue dots with an
      opacity reflecting their respective weights in the fit. Middle panel:
      Calibration residuals $r=m_{\mathrm{SMP}} - m_{\mathrm{PS1}} - \alpha
      \times \mathrm{col} - zp_{\mathrm{SMP}}$ as a function of the color. The
      vertical lines shows the applied color cut. The horizontal black lines
      show zero. Bottom panel: Calibration residuals as a function of the PS1
      magnitudes.}
    \label{fig:dr2_lccalib}
  \end{center}
\end{figure*}

All produced light curves have been calibrated against the PS1 catalog, as
described in \Sect~\ref{subsec:smp_pipeline:smp_lightcurve_calibration}, using
the scene modeling fluxes (averaged from the scene modeling light curves) as
anchors. The calibration model reads
\begin{equation}
  \hat{m}_{\mathrm{SMP},i} = m_{\mathrm{cal},i} + \alpha \times \mathrm{col} + \delta zp(\vec{x}) + zp_{\mathrm{SMP}},
\end{equation}
where $\hat{m}_{\mathrm{SMP},i} = -2.5\log_{10} \hat{f}_{\mathrm{SMP},i}$ and
$m_{\mathrm{cal},i}$ are the instrumental and calibrated magnitudes of field
star $i$, respectively. The term $\alpha\ \times\ \mathrm{col}$ accounts for the
filter differences between ZTF and PS1, while $zp_{\mathrm{SMP}}$ is the
calibration zero-point applied to the light curve --the only parameter of
interest. The calibration zero point is registered along with the light curve
points in the final output.

Figure \ref{fig:dr2_lccalib} shows representative calibration residuals for a
subset of the light curves, across the three ZTF bands: $g$, $r$, and $i$. In
the top panel, calibration residuals $m_{\mathrm{SMP}} - m_{\mathrm{PS1}}$ are
plotted as a function of the PS1 $g-i$ color, highlighting noticeable color
terms between the ZTF and PS1 band-passes. These slopes indicate that the $g$,
$r$ and $i$ ZTF band-passes are respectively \SI{6}{\nano\meter} bluer,
\SI{14}{\nano\meter} redder and \SI{15}{\nano\meter} redder than the PS1
band-passes. The middle panel displays the color-corrected calibration
residuals. As can be seen, the PS1/ZTF band-pass differences are large enough to
induce slight nonlinearities in the color transformations, which becomes
negligible when the color range of calibration stars is reduced. Finally, the
bottom panel of \Fig \ref{fig:dr2_lccalib} presents the calibration residuals as
a function of magnitude, revealing significant residual nonlinearities ranging
from 10 to 20 mmag across the \SIrange{14}{18}{\mag} range.

Disentangling the nonlinearity sources is a nontrivial task as some fraction
of it may also come from the PS1 calibration catalog. However, ongoing work on
our internal calibration catalog \CitepInPrep{Racine}, based on aperture
photometry and verified to be immune to the pocket effect, shows a residual
nonlinearity of the same order of magnitude for faint stars. Thus, we attribute
with high confidence that these residual nonlinearities origin from the pocket
effect.
\section{Comparison with forced photometry}\label{sec:comparison_forced_photometry}

\begin{figure}[t]
  \begin{center}
    \includegraphics[width=0.8\linewidth]{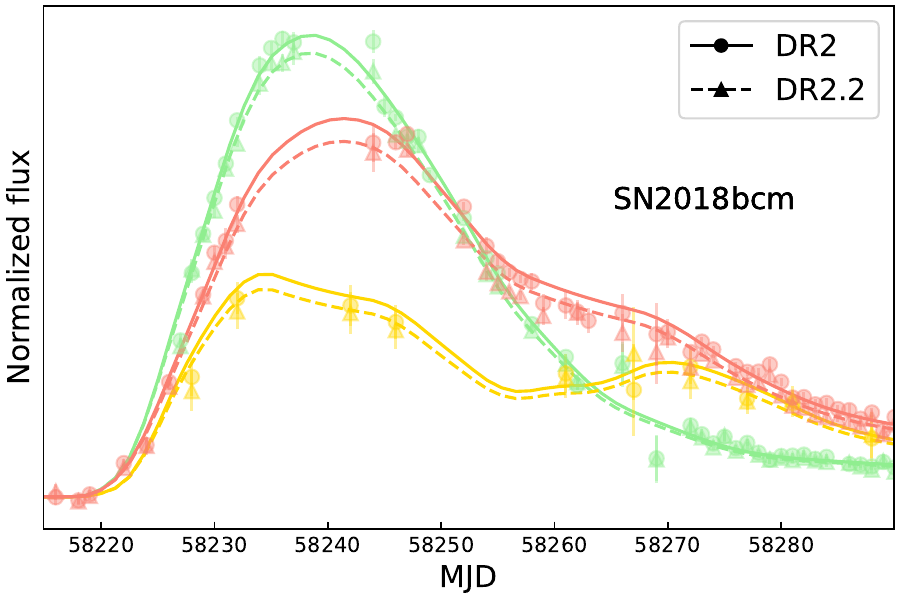}
    \includegraphics[width=0.8\linewidth]{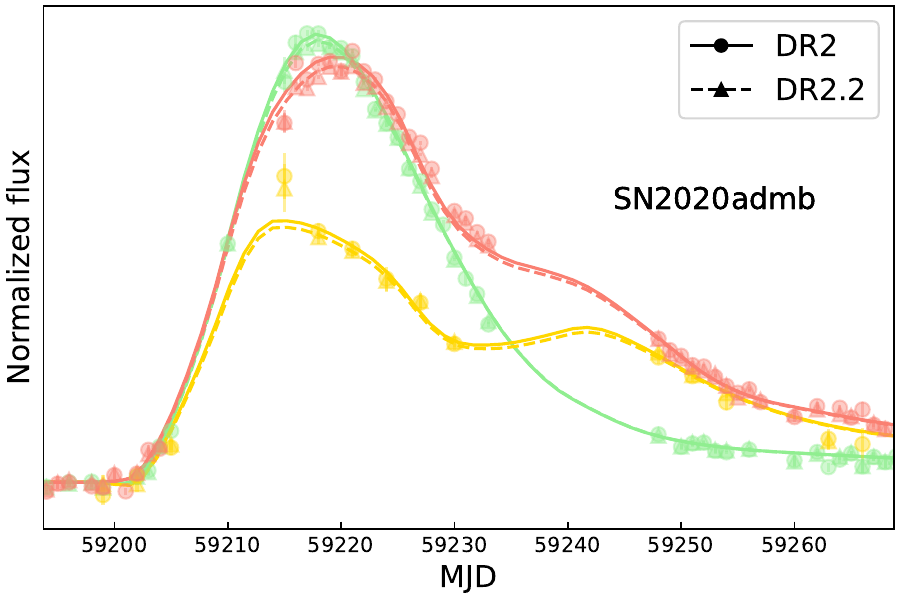}
    \caption{Comparison of the forced and SMP for two SNe: SN2018bcm
      (ZTF18aakzliv) and SN2020admb (ZTF20adadsgm). The $g$, $r$, and $i$ bands are
       shown in green, red, and orange, respectively. Circles--lines and
      triangles--dashed lines represent respectively the forced and the SMP and
      their respective SALT2 models.}
    \label{fig:lc_compa}
  \end{center}
\end{figure}

We now compare our SMP flux measurements with the recently released DR2
measurements from \CiteInPrep{Smith}. Figure \ref{fig:lc_compa} presents two
typical SMP light curves from our internal release, alongside the corresponding
DR2 forced photometry light curves. Overall the photometry methods show good
agreement, although forced photometry fluxes appear slightly brighter than those
obtained from scene modeling.

\begin{figure}[t]
  \begin{center}
    \includegraphics[width=\linewidth]{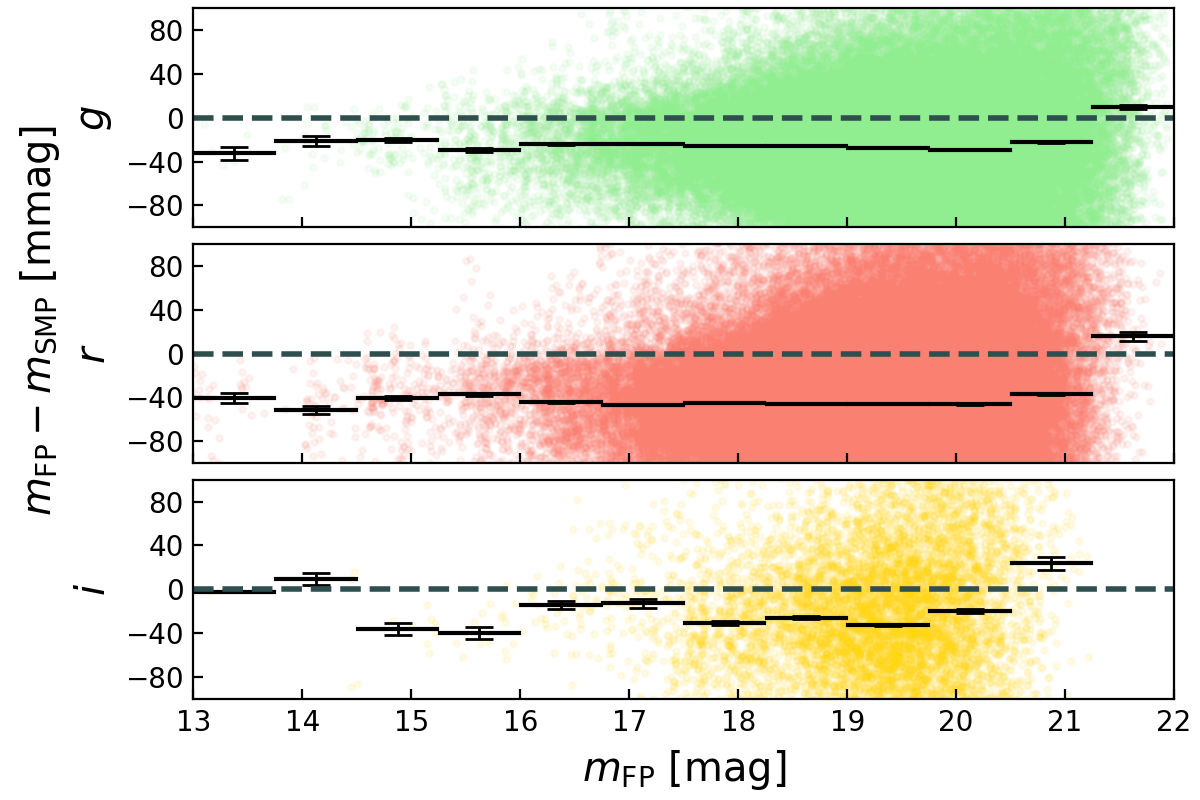}
    \caption{Comparison between DR2 forced photometry (FP) and SMP magnitudes.
      The $g$, $r$, and $i$ bands are shown in green (top panel), red (middle
      panel), and orange (bottom panel). The black segments represent the binned
      data where the $y$-axis error is the bin uncertainty. The gray dotted
      lines show zero. We note that we selected only the photometric points with
      an error $<$ \SI{0.3}{\mag}.}
    \label{fig:dr2_vs_dr22}
  \end{center}
\end{figure}

To further investigate these discrepancies, Fig. \ref{fig:dr2_vs_dr22} shows a
systematic comparison between both photometries. Although photometry from both
sources aligns reasonably well across all three bands, systematic differences
are evident across the flux estimates. The largest differences are observed in
the $g$ and $r$ bands, of the order of 30 and 50~mmag respectively. In the $i$
band, the average difference is only of \SI{20}{\milli\mag}. In all three bands,
objects are, on average, brighter in DR2. From the binned data (black dots), no
evolution is observed with respect to magnitude; brighter and fainter objects
show the same offset in magnitudes between the two photometry sets. Identifying
the exact source of these differences is challenging, as DR2 photometry lacks
light curves for field stars. In contrast, the SMP calibration chain is better
understood, since it relies explicitly on the SMP light curves of the field
stars. The observed offset between the forced photometry-based DR2 and SMP
fluxes is compatible with the photometric calibration uncertainty affecting the
DR2 fluxes -- \SI{30}{\milli\mag}, as reported in \CiteInPrep{Smith}. This level
of uncertainty is substantially higher than the mmag-level accuracy needed for
these light curves to be suitable for cosmological measurements.

\begin{figure}[t]
  \begin{center}
    \includegraphics[width=\linewidth]{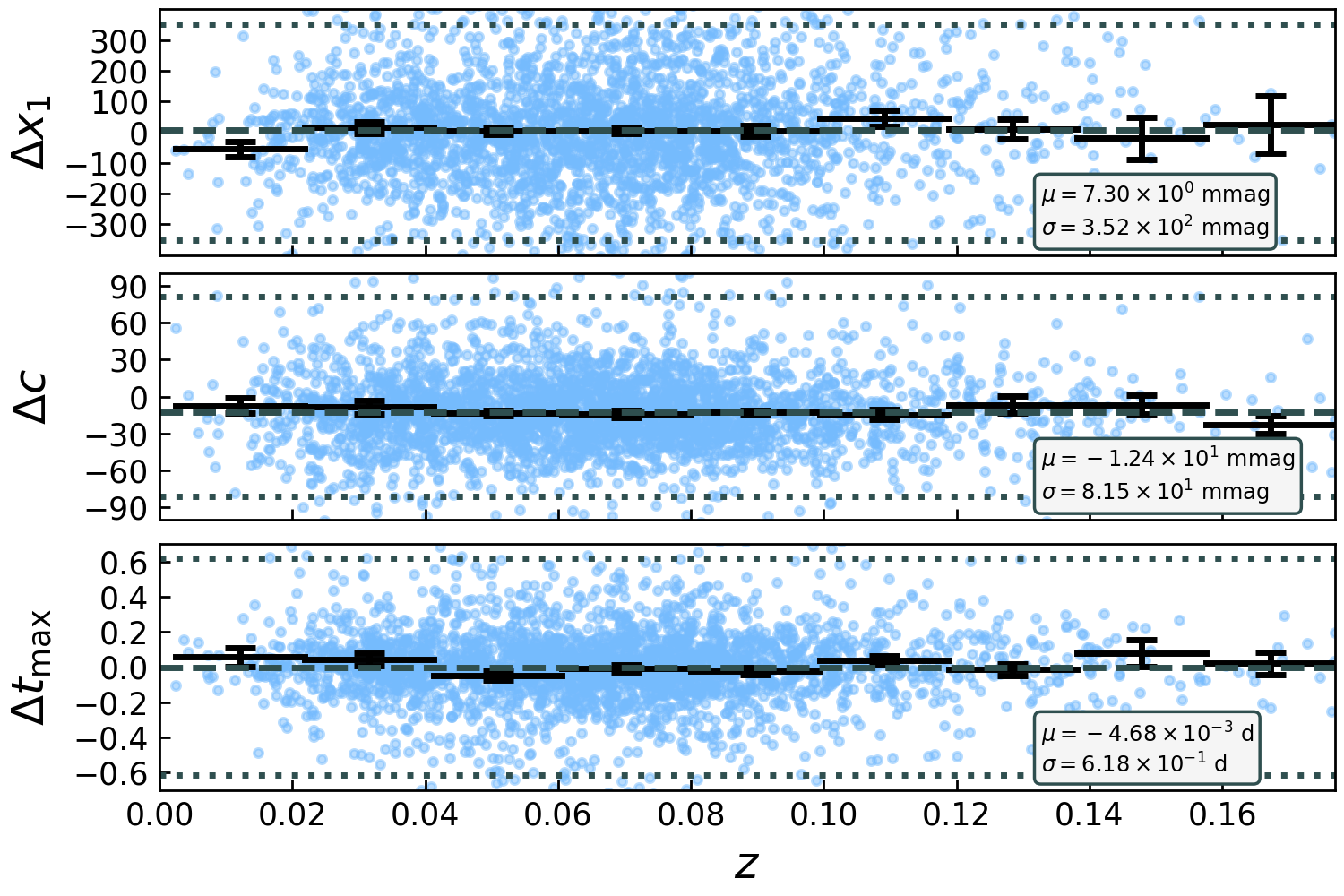}
    \caption{Comparison of the $x_{1}$, color and $t_{\mathrm{max}}$ derived
      from the DR2 forced photometry (FP) and SMP processings using the SALT2.4
      light curve fitter. The densely dotted gray lines show the mean value, and
      the more sparsely dotted lines the $\pm 1$ standard deviation interval. As
      shown, the two sets of estimates are consistent on average. }
    \label{fig:x1ctmax_dr2_dr22}
  \end{center}
\end{figure}

To derive the distance moduli and light curve parameters, we use the Tripp
relation \citep{trippTwoparameterLuminosityCorrection1998}
\begin{align} \mu = -2.5 \log_{10}(x_0) + \alpha x_1 - \beta c - M_B, \end{align}
where $x_0$, $x_1$, and $c$ are, respectively, a simple multiplicative scaling
factor, the stretch, and the color. These fitted parameters together with the
time of maximum brightness ($t_{\mathrm{max}}$) are obtained by adjusting our
light curves using
SNCosmo\footnote{\url{https://sncosmo.readthedocs.io/en/stable/about.html}} with
the SALT2 model \citep{guySALT2UsingDistant2007} version 4
\citep{betouleImprovedCosmologicalConstraints2014}, hereby noted SALT2.4. The
values of $\alpha$, $\beta$, and $M_B$ are taken from
\citet{betouleImprovedCosmologicalConstraints2014}.

In Fig. \ref{fig:x1ctmax_dr2_dr22} we compare the $x_{1}$, $c$ and
$t_{\mathrm{max}}$ parameters derived from the DR2 data release and our SMP
internal data release using the SALT2.4 light curve fitter. As expected from the
direct comparison of the SN magnitudes, we observe a small, redshift-independent
offset of \SI{10}{\milli\mag} in color. The $x_{1}$ and $T_{max}$ estimates are
fully consistent between the two datasets. We therefore conclude that the
discrepancies between DR2 and SMP discussed in this section do not have a
significant impact on the DR2 papers discussing SNe Ia populations.

\begin{figure}[t]
  \begin{center}
    \includegraphics[width=\linewidth]{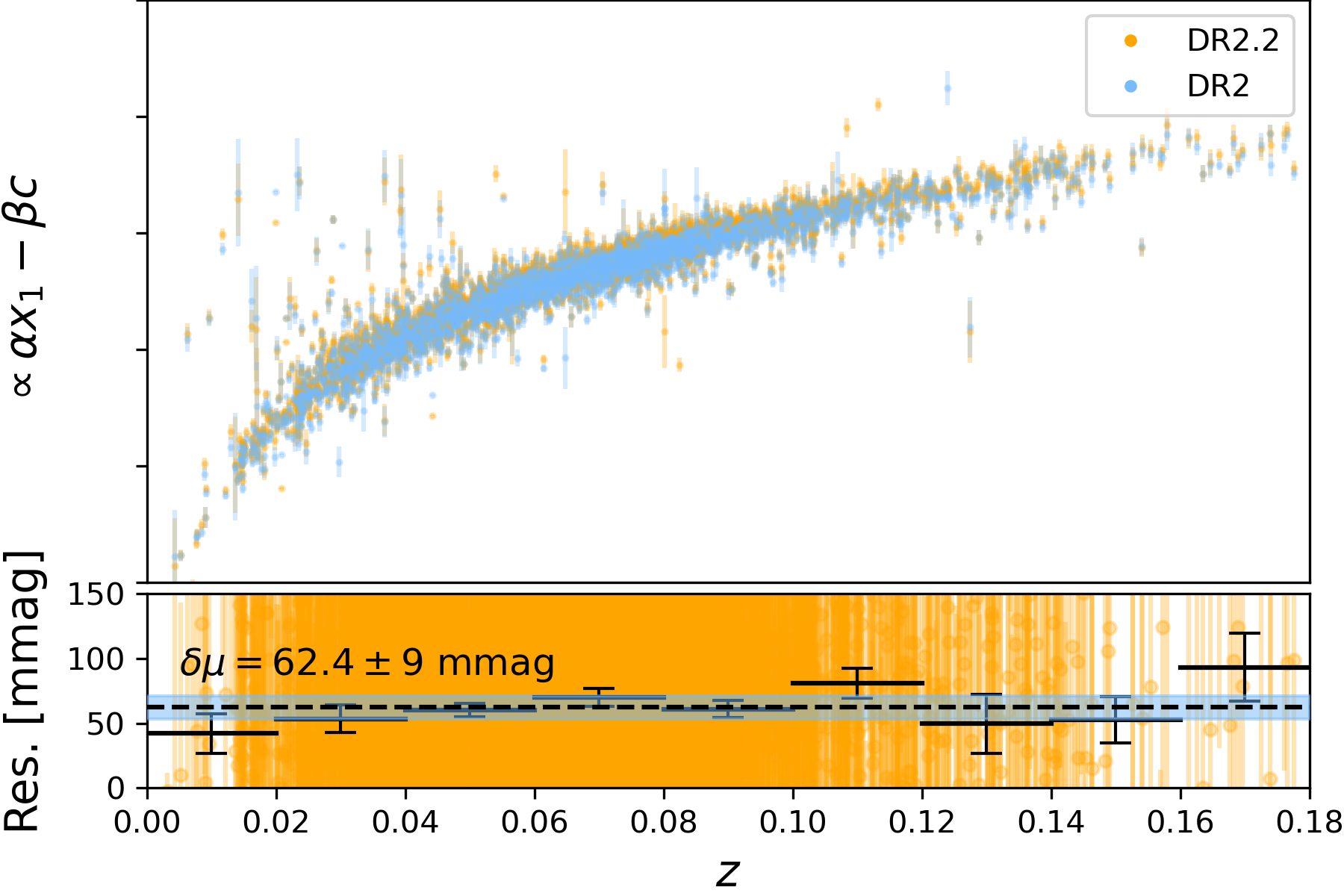}
    \caption{Hubble diagram for the forced photometry (blue squares) and the
      scene modeling magnitudes (orange dots). The residuals are plotted in the
      bottom panel together with the binned data (black dots). The weighted
      average residual is shown as a horizontal dotted line, with the blue
      region marking the $\pm$1 standard deviation interval. We note that the bin
      $y$-axis errors are the bin uncertainties.}
      \label{fig:hd_dr2_dr22}
  \end{center}
\end{figure}

Finally, we try to assess the impact of these systematic differences in terms of
distances. In Fig. \ref{fig:hd_dr2_dr22}, we present the Hubble diagram derived
from both photometry dataset sets.

As shown in Fig. \ref{fig:hd_dr2_dr22}, the distance moduli from both photometry
sets differ by about \SI{62}{\milli\mag}. The SN distances from the SMP
photometry are slightly larger than those obtained from the DR2 photometry. This
is consistent with \Fig \ref{fig:dr2_vs_dr22}, where we observe that the DR2
photometry is, on average, brighter than the SMP photometry
($\sim$\SI{50}{\milli\mag}). Furthermore, a slight variation in redshift can be
noted. This result is compatible with the 5\% accuracy level reported
in~\cite{rigaultZTFSNIa2025a} for the DR2 forced photometry.

The discrepancy between the DR2 forced photometry and SMP discussed in this work
is significantly larger than the level of precision we aim to achieve. It is
important to note, however, that the forced photometry used in DR2 was not
designed to reach sub-percent accuracy. Rather, it was intended as an efficient
and computationally inexpensive approach to rapidly produce an early version of
their light curves. To this end, several approximations were introduced to
accelerate the processing. For instance, a nonvariable PSF model is used to
estimate fluxes on the difference images, where ZOGY advises subdividing the
images into subquadrants, each with their own PSF model to account for spatial
variability \citep{zackayProperImageSubtraction2016}. In addition, the
astrometric solution is not recomputed and instead relies on the WCS provided by
the exposures. These approximations inevitably impact the resulting photometry.
Accordingly, \cite{rigaultZTFSNIa2025a} reports a 5\% uncertainty, consistent
with the offset we measure here.

As discussed in \Sect~\ref{sec:sn_ia_light_curves}, forced photometry also
presents intrinsic limitations that make it difficult to identify the exact
origin of these discrepancies. In particular, by construction, forced photometry
does not provide measurements for field stars. As a consequence, it is not
possible to directly verify the calibration of the difference images on which
the photometry is performed, nor explicitly assess how the aforementioned
approximations propagate into the flux estimates.

The upcoming DR2.5 release will include a dedicated study based on realistic
simulations to investigate how SMP and forced photometry preserve flux.

\section{Limitations}\label{sec:limitations}
We now examine the specific limitations of the current processing that have
rendered our data unsuitable for cosmological analyses. These limitations are
not unique to our data release; they impact both the DR2 forced-photometry and
the SMP presented in this paper.

The first of the three limiting factors is the brighter-fatter effect
\citep[][]{antilogusBrighterfatterEffectPixel2014,guyonnetEvidenceSelfinteractionCharge2015}.
This sensor effect makes the PSF of the brighter stars wider than that of the
fainter stars. It has recently been measured as being significative and we are
currently implementing its correction as presented in
\cite{astierCorrectionBrighterfatterEffect2023}. The second limitation is a
novel sensor effect that distorts the PSF of the faint stars relative to that of
the brighter stars that we have dubbed the pocket effect. The third and last
limitation we discuss here arises from the insufficient knowledge of the
band-passes throughput.

\subsection{Pocket effect}\label{subsec:limitations:sensor_effects}
\begin{figure}[t]
  \begin{center}
    \includegraphics[width=\linewidth]{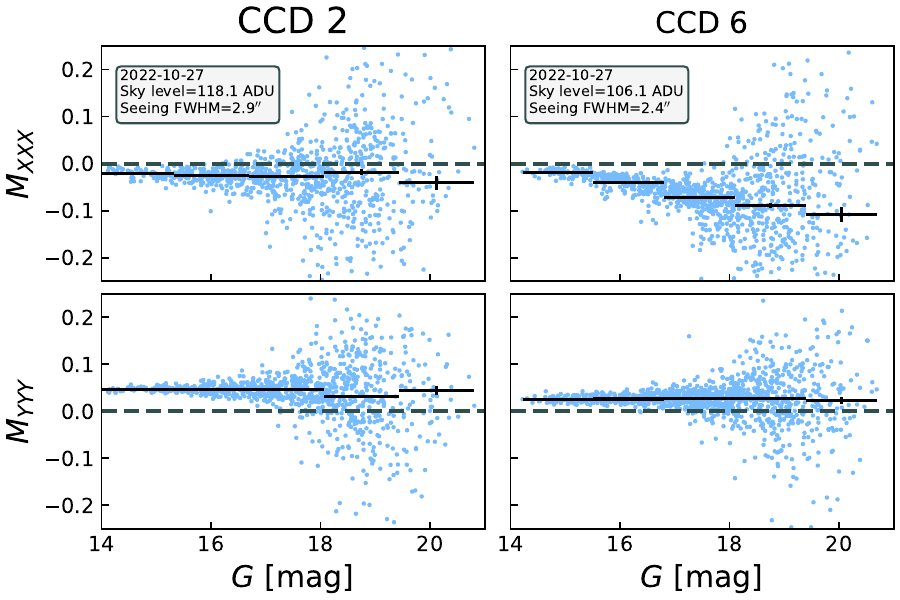}
    \caption{PSF skewness as a function of $G$ magnitude for CCDs 2 and 6 from a
      single exposure. The horizontal gray dotted lines show zero. A pronounced
      skewness with respect to magnitude is observed on the $x$-axis of CCD~6,
      indicating a significant magnitude-dependent asymmetry in the PSF for this
      particular detector.}
    \label{fig:skewness_ccd}
  \end{center}
\end{figure}

As shown in Fig. \ref{fig:astrometric_residuals} (in
\Sect~\ref{subsec:smp_pipeline:astrometric_transforms}), the astrometric
residuals in the serial-direction exhibit a noticeable dependence on flux, while
the residuals in the parallel direction behave normally. This effect was traced
back to a flux-dependent distortion, centroid shift, of the PSF shape as
illustrated in Fig. \ref{fig:skewness_ccd}. Specifically, we observe that the
skewness of the PSF in the $x$-direction, as quantified by the third order
moment $M_{xxx}$ varies significantly with flux. In contrast, the equivalent
third order moment in the parallel direction $M_{yyy}$ remains constant as a
function of flux.

These PSF distortions appear to uniformly affect all objects within the same
quadrant. The intensity of the effect varies significantly from CCD-to-CCD, with
some sensors being nearly unaffected while most are impacted to varying degrees.
Additionally, slight variations in intensity are observed from quadrant to
quadrant within the same sensor. The issue was traced to be a readout artifact.

Although the effect is detectable throughout the entire survey, it is less
pronounced (close to insignificant) before November 2019, when an upgrade of the
readout waveforms has been made. We finally note that the amplitude of the
effect is influenced by the CCD illumination level, with its intensity peaking
when the sky background is low; it becomes almost non-detectable beyond
approximately 6000~ADU per pixel of sky background.

The most plausible physical explanation is that a fraction of the charges can
escape into a pocket during pixel readout. The exact nature of this pocket
remains uncertain: it could be a classical trap, albeit with a large capacity of
a few hundred electrons, or a defect in the drift field. These charges are
subsequently retrieved after a relatively long delay, typically on the order of
a few dozen microseconds. The likelihood of charge escape depends on the
pocket's content: it is low when the pocket is full (especially when the sky
background is high) and significantly higher when the pocket is empty.

This issue, now identified as the pocket effect, has long been visible in ZTF
data, especially in CCD-06. An empirical stellar PSF-flux variability correction
method, called
Zubercal,\footnote{\url{http://atua.caltech.edu/ZTF/Zubercal.html}} has
been developed to patch science image-based light curves from visible
detector effects and does somewhat absorbs the pocket effect. However, these light
curves are not those obtained from difference images, and are thus limited to
 stellar objects or spatially isolated transients, for example. There are no Zubercal
equivalents for alert or force-photometry services, and Zubercal itself is not in
production for the publicly available ZTF science image light curve service.
Consequently, the pocket effect affects all transient light curves from ZTF
starting November 2019.

\subsection{Impacts of the pocket effect}\label{subsec:limitations:impact_pocket_effect}

\paragraph{Linearity:} The flux-dependent distortions of the PSF shape are not
accounted for by the current PSF model used by the scene modeling algorithm.
This results in nonlinearities of the SMP flux estimator ranging from 1--2\% on
most sensors, and up to 6--7\% on the most affected CCDs.

\begin{figure}[t]
\begin{center}

\begin{subfigure}{0.85\linewidth}
\includegraphics[width=\textwidth]{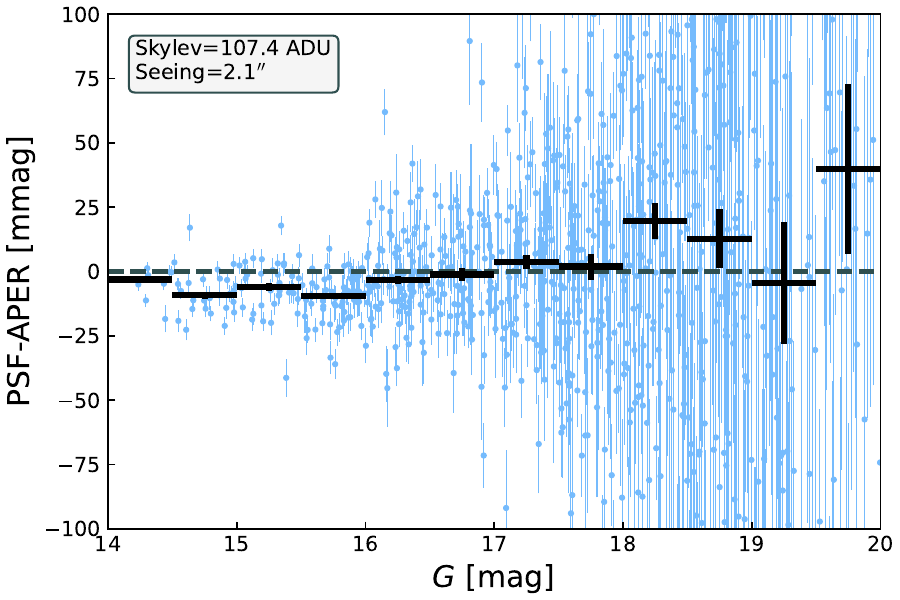}
\caption{CCD 2 -- Q1}
\end{subfigure}

\bigskip

\begin{subfigure}{0.85\linewidth}
\includegraphics[width=\textwidth]{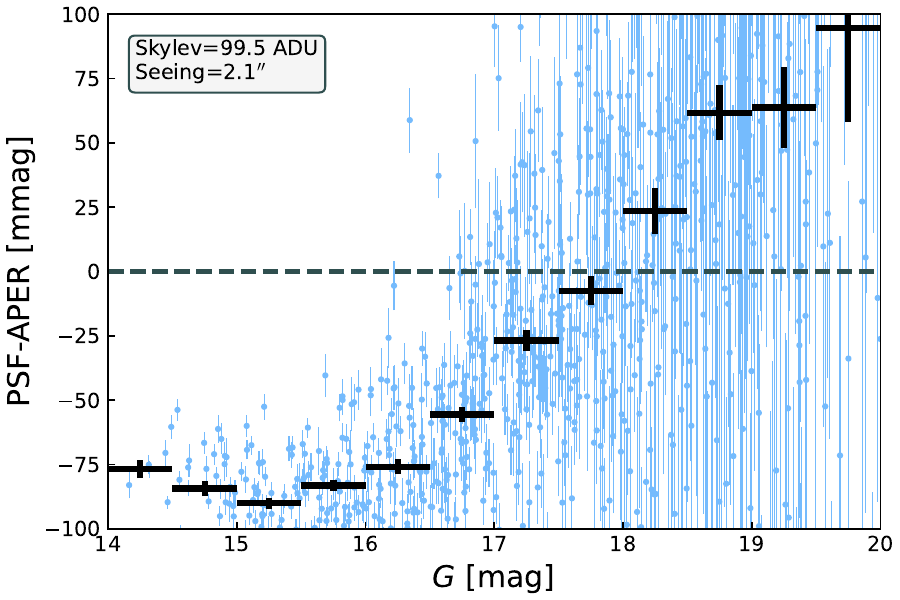}
\caption{CCD 6 -- Q1}
\end{subfigure}

\bigskip

\begin{subfigure}{0.85\linewidth}
\includegraphics[width=\textwidth]{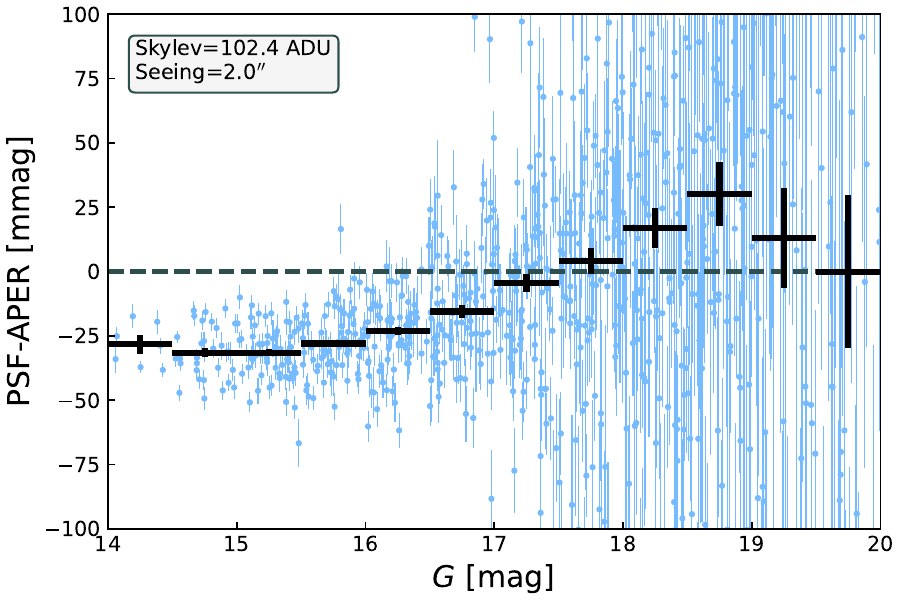}
\caption{CCD 14 -- Q1}
\end{subfigure}
    \caption{Upper panel: $m_{\mathrm{PSF}}-m_{\mathrm{aper}}$ as a function of
      the \textit{Gaia} $G$-magnitude on a low background frame taken with
      CCD~2, showing negligible nonlinearities. Center panel: Same exposure on
      CCD~6, which shows nonlinearities of about 8\% peak-to-peak. Lower panel:
      Same exposure on CCD~14. Although this sensor is less affected by the
      pocket effect, nonlinearities at the level of 2\% (peak-to-peak) can still
      be observed. The horizontal gray dotted lines show zero.}
\label{fig:limitations:psf_vs_aper}
\end{center}
\end{figure}

To illustrate this, we compare in Fig. \ref{fig:limitations:psf_vs_aper} the PSF
and aperture fluxes measured on two ZTF quadrants. At first order, aperture
photometry is not sensitive to the fine details of the PSF, hence, not sensitive
to the pocket effect. As can be seen, on low background exposures, we observe
large nonlinearities, ranging from 1\% to 7\% affecting the PSF-to-aperture
photometry ratio. Again, this is well above our accuracy target of 0.1\%. This
invalidates the assumption of PSF homothety underlying PSF modeling techniques
and photometry. As a result, nonlinearities arise in the PSF flux estimates,
reaching up to $\sim$10\% (peak-to-peak) on the most affected sensors, over the
magnitude range 14--20. This alter the shape of the SN light curve and
compromises accurate calibration transfer from bright calibrators to the
faint supernova.

\paragraph{Astrometry:}
Furthermore, these PSF distortions (centroid shifts) cause flux-dependent
distortions in astrometry, as illustrated in
Fig. \ref{fig:astrometric_residuals}. Since the scene modeling fit enforces
relative source positions using pre-determined astrometric transformations
(\Sect~\ref{subsec:smp_pipeline:astrometric_transforms}), this introduces an
additional flux-dependent bias, underestimating the SN flux.

Due to the extreme dependence of these nonlinearities on observing conditions
and its impact on photometry and astrometry, we have chosen to develop a
pixel-level correction, applied on each frame during detrending. This correction
and its validation will be discussed in \CiteInPrep{Regnault}.

\subsection{Band-pass models}\label{subsec:limitations:bandpass_models}
\begin{figure}[h]
  \begin{center}
    \includegraphics[width=\linewidth]{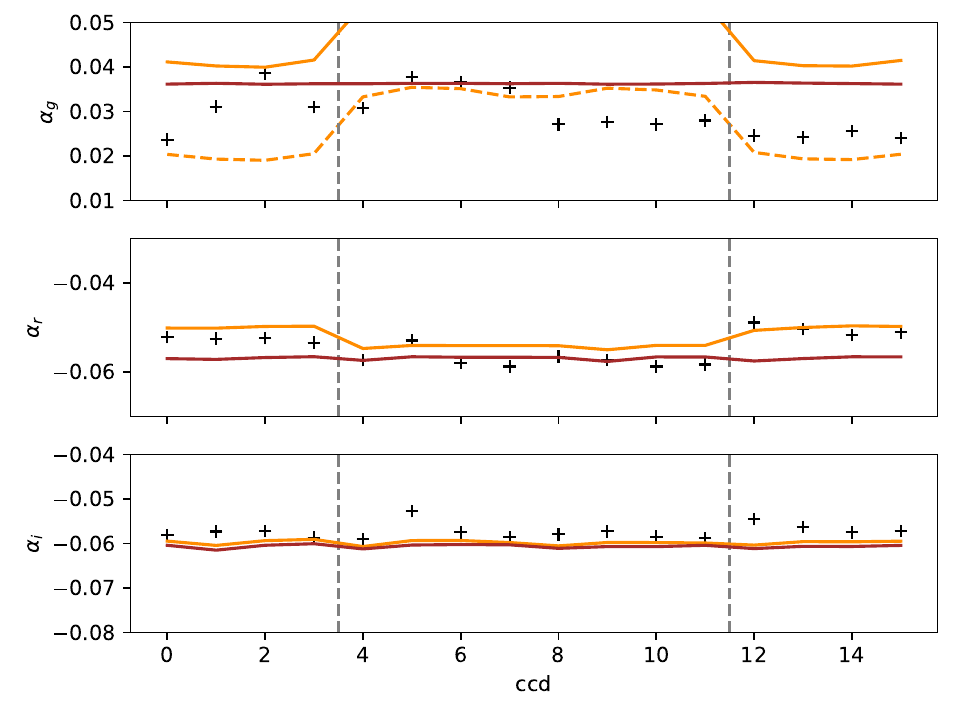}
    \caption{Color transformation coefficients $\alpha$ between ZTF and PS1 as a
      function of CCD number. The values obtained from stellar measurements are
      shown in black, and are compared to predictions built from synthetic
      photometry using \textit{Gaia} spectra (orange and red lines). The red line
      corresponds to predictions using the original \texttt{SNCosmo} band-passes;
      the orange line uses spatially variable band-passes derived from bench
      measurements provided by the ZTF team. In the upper panel ($g$ band), the
      dotted orange line uses the same spatially variable band-pass model,
      shifted by \SI{3}{\nano\meter}.}
    \label{fig:synthetic_color}
  \end{center}
\end{figure}

\subsection{Band-pass models}
Band-pass models are a crucial ingredient of any SN cosmology analysis. They are
used in the survey calibration stage, to derive the survey zero points in each
band from the Primary Flux Standard spectra. They are also used in the SN
distance estimation process, when deriving the SN restframe flux and color from
the SN observer frame light curves using a spectrophotometric model. At first
order, passband uncertainties can be quantified as uncertainties on the passband
average position in wavelength ($\bar{\lambda} = \int \lambda T(\lambda)
d\lambda / \int T(\lambda) d\lambda$). For the measurement of the dark energy
EoS (supplementing this dataset with higher redshift data) the
requirement is that the accuracy on $\bar{\lambda}$ should be of the order of
$\sim$\SI{3}{\angstrom}.

A straightforward first-order approach to assess the accuracy of band-pass
models is to compare the empirically measured color-terms between a survey
telescope and a similar instrument with known band-passes to those predicted
using synthetic photometry. In Fig. \ref{fig:synthetic_color} we show the slopes
of the calibration color-color plots between ZTF and PS1, for the $g$, $r$, and
$i$ bands, measured from the observational data (black points) and that
predicted using synthetic photometry using the \texttt{SNCosmo} ZTF and PS1
bandbass models (red lines), plotted as a function of CCD number. A distinct
difference is visible between the single coating CCDs (0--3, 12--15) and double
coating CCDs (4--11). It corresponds to a color-term difference of about 0.01
(\resp~0.05) in the $g$ (\resp $r$) bands, equating to band-pass differences of
about \SI{30}{\angstrom} (\resp~\SI{15}{\angstrom}).

The color terms derived from synthetic photometry using the PS1 and ZTF
band-passes published in \texttt{SNCosmo} are indicated with a red line. The ZTF
band-passes distributed through \texttt{SNCosmo} are actually an average between
the band-pass models built for double-coating and single-coating sensors, and
fail to capture the difference between these two CCD types. As a consequence,
the \texttt{SNCosmo} ZTF band-passes\footnote{labeled \texttt{ztfg},
\texttt{ztfr}, \texttt{ztfi}} are inaccurate by about $\sim$\SI{30}{\angstrom}
in $g$ and \SI{15}{\angstrom} in $r$ across approximately half of the focal
plane.

Efforts are underway to develop a more precise model of the ZTF band-passes,
using laboratory scans of the transmissions of the various elements of the ZTF
optical system. The synthetic color terms generated with this improved model are
shown in orange. As seen, these models accurately reproduce data in the $r$
band, but deviate substantially (by roughly \SI{60}{\angstrom}) in the $g$ band.
We conclude that an adequately accurate model of the ZTF band-passes fulfilling
our accuracy requirements is not yet available.

\section{Conclusion}
\label{sec:conclusion}
We have shown that the ZTF scene modeling pipeline is capable of efficiently
handling the extensive data volumes delivered by the ZTF observing system. The
pipeline can handle the 3.2~millions frames, amounting to approximately
\SI{179}{\tera\byte}, of the DR2 dataset in two to three weeks, facilitating
multiple iterations across the entire dataset. With specific adjustments on
identified critical sections of the code (notably the PSF modeling code), we
anticipate that it could process the full ZTF spectroscopic dataset (comprising
9,000 SNe~Ia) within a month. With a re-organization and further optimizations,
the pipeline could potentially process the entire ZTF photometric dataset
(30,000 objects) in less than six months.

The pipeline generates high-quality light curves for the SNe and their
surrounding field stars using a consistent PSF flux estimator, effectively
reducing biases caused by systematic differences between flux estimators.
Furthermore, with most primary flux standard candidates inherently part of the
ZTF dataset, we can implement a concise, redundant, and robust photometric
calibration chain that directly connects SN fluxes to the calibration sources.
This design ensures that the survey and analysis chain are robust, enabling us
to aim for 0.1\% accuracy in SN photometry.

Nevertheless, significant challenges remain before the scene modeling light
curves can be utilized for a cosmological analysis:
\begin{enumerate}
  \item In addition to a non-negligible amount of brighter--fatter, we have
    identified a new sensor effect, which we have called the pocket effect,
    which distorts the PSF shape as a function of flux. The combination of the two
    effects introduces substantial  nonlinearities (up to 7\%) in the PSF flux
    estimates. The pocket effect also introduces differential biases in the
    measured position of bright  and faint sources  (\eg, field stars vs. SNe).
  \item While the DR2 and SMP determinations of $x_{1}$ and color are
    consistent, we have uncovered a significant discrepancy between the DR2 and
    SMP fluxes in the $g$ and $r$ bands. Although the SMP calibration chain is
    robust (since the SMP photometry can be directly applied to the surrounding
    field stars), we cannot rule out the possibility that the large
    nonlinearities identified in this study contribute to this
    discrepancy.
  \item The accurate estimation of SN distances requires precise models of the
    instrument band-passes. The current models are precise at the 3 nm level,
    which remains about an order of magnitude above our requirements.
\end{enumerate}

Given these unsolved systematics, we conclude that neither the current DR2
fluxes nor the SMP measurements can be used for precise cosmological analyses.

To address these issues, we have outlined a comprehensive plan. First, a
pixel-level correction has been developed to mitigate the pocket effect and has
been fully integrated into a revised detrending process. Second, a redundant
calibration chain connecting the CALSPEC library and the SMP light curves is
being developed. It implements two routes, one based on a full sky ZTF catalog,
and another based on the \textit{Gaia} spectrophotometry. Finally, new band-pass
models, constructed from bench measurements of all optical elements provided by
the Caltech team, are currently being validated using stellar observations and
will be released as part of the SNCosmo framework. Additionally, an in situ
measurement campaign of the ZTF filters, using a collimated beam projector
\citep[][]{neveuAccurateMeasurementTelescope2024}, is in preparation to further
validate and refine these filter models.

These are critical steps toward preparing the data for future analyses. The
forthcoming DR2.5 release is expected to address these issues, and to provide a
reliable dataset for cosmological studies. This paper concludes the DR2 series,
and opens the way to the upcoming DR2.5 release.

\begin{acknowledgements}
This work has been supported by the Agence Nationale de la Recherche of the
French government through the program ANR-21-CE31-0016-03.

This project has received funding from the European Research Council (ERC)
under the European Union’s Horizon 2020 research and innovation program (grant
agreement n759194-USNAC).

Based on observations obtained with the Samuel Oschin Telescope 48-inch and the
60-inch Telescope at the Palomar Observatory as part of the \textit{Zwicky} Transient
Facility project. ZTF is supported by the National Science Foundation under
Grants No. AST-1440341 and AST-2034437 and a collaboration including current
partners Caltech, IPAC, the Weizmann Institute of Science, the Oskar Klein
Center at Stockholm University, the University of Maryland, Deutsches
Elektronen-Synchrotron and Humboldt University, the TANGO Consortium of Taiwan,
the University of Wisconsin at Milwaukee, Trinity College Dublin, Lawrence
Livermore National Laboratories, IN2P3, University of Warwick, Ruhr University
Bochum, Northwestern University and former partners the University of
Washington, Los Alamos National Laboratories, and Lawrence Berkeley National
Laboratories. Operations are conducted by COO, IPAC, and UW.

SED Machine is based upon work supported by the National Science Foundation
under Grant No. 1106171.

This work has made use of data from the European Space Agency (ESA) mission \textit{Gaia}
(\url{https://www.cosmos.esa.int/gaia}), processed by the \textit{Gaia} Data Processing
and Analysis Consortium (DPAC,
\url{https://www.cosmos.esa.int/web/gaia/dpac/consortium}). Funding for the DPAC
has been provided by national institutions, in particular the institutions
participating in the \textit{Gaia} Multilateral Agreement.

T.E.M.B. is funded by Horizon Europe ERC grant no. 101125877.

U.B. is funded by Horizon Europe ERC grant no. 101125877.

This work has been supported by the research project grant “Understanding the
Dynamic Universe” funded by the Knut and Alice Wallenberg Foundation under Dnr
KAW 2018.0067 and the {\em Vetenskapsr\aa det}, the Swedish Research Council,
project 2020-03444.

JHT is funded by Horizon Europe ERC grant no. 101125877.

KM is supported by Horizon Europe ERC grant no. 101125877.

L.G. acknowledges financial support from AGAUR, CSIC, MCIN and AEI
10.13039/501100011033 under projects PID2023-151307NB-I00, PIE 20215AT016,
CEX2020-001058-M, ILINK23001, COOPB2304, and 2021-SGR-01270.
\end{acknowledgements}

\bibliographystyle{aa}
\bibliography{aanda}

@article{abdulkarimDESIDR2Results2025,
  title = {{{DESI DR2}} Results. {{II}}. {{Measurements}} of Baryon Acoustic Oscillations and Cosmological Constraints},
  author = {Abdul Karim, M. and Aguilar, J. and Ahlen, S. and Alam, S. and Allen, L. and Allende Prieto, C. and Alves, O. and Anand, A. and Andrade, U. and Armengaud, E. and Aviles, A. and Bailey, S. and Baltay, C. and Bansal, P. and Bault, A. and Behera, J. and BenZvi, S. and Bianchi, D. and Blake, C. and Brieden, S. and Brodzeller, A. and Brooks, D. and {Buckley-Geer}, E. and Burtin, E. and Calderon, R. and Canning, R. and Rosell, A. Carnero and Carrilho, P. and Casas, L. and Castander, F. J. and Charles, M. and Chaussidon, E. and {Chaves-Montero}, J. and Chebat, D. and Chen, X. and Claybaugh, T. and Cole, S. and Cooper, A. P. and Cuceu, A. and Dawson, K. S. and {de la Macorra}, A. and {de Mattia}, A. and Deiosso, N. and Della Costa, J. and Demina, R. and Dey, A. and Dey, B. and Ding, Z. and Doel, P. and Edelstein, J. and Eisenstein, D. J. and Elbers, W. and Fagrelius, P. and Fanning, K. and {Fern{\'a}ndez-Garc{\'i}a}, E. and Ferraro, S. and {Font-Ribera}, A. and {Forero-Romero}, J. E. and Frenk, C. S. and {Garcia-Quintero}, C. and Garrison, L. H. and Gazta{\~n}aga, E. and {Gil-Mar{\'i}n}, H. and Gontcho A Gontcho, S. and Gonzalez, D. and {Gonzalez-Morales}, A. X. and Gordon, C. and Green, D. and Gutierrez, G. and Guy, J. and Hadzhiyska, B. and Hahn, C. and He, S. and Herbold, M. and {Herrera-Alcantar}, H. K. and Ho, M.-F. and Honscheid, K. and Howlett, C. and Huterer, D. and Ishak, M. and Juneau, S. and Kamble, N. V. and Kara{\c c}ayl{\i}, N. G. and Kehoe, R. and Kent, S. and Kim, A. G. and Kirkby, D. and Kisner, T. and Koposov, S. E. and Kremin, A. and Krolewski, A. and Lahav, O. and Lamman, C. and Landriau, M. and Lang, D. and Lasker, J. and Le Goff, J. M. and Le Guillou, L. and Leauthaud, A. and Levi, M. E. and Li, Q. and Li, T. S. and Lodha, K. and Lokken, M. and {Lozano-Rodr{\'i}guez}, F. and Magneville, C. and Manera, M. and Martini, P. and Matthewson, W. L. and Meisner, A. and {Mena-Fern{\'a}ndez}, J. and Menegas, A. and Mergulh{\~a}o, T. and Miquel, R. and Moustakas, J. and {Mu{\~n}oz-Guti{\'e}rrez}, A. and {Mu{\~n}oz-Santos}, D. and Myers, A. D. and Nadathur, S. and Naidoo, K. and Napolitano, L. and Newman, J. A. and Niz, G. and Noriega, H. E. and Paillas, E. and {Palanque-Delabrouille}, N. and Pan, J. and Peacock, J. A. and Pellejero Ibanez, M. and Percival, W. J. and {P{\'e}rez-Fern{\'a}ndez}, A. and {P{\'e}rez-R{\`a}fols}, I. and Pieri, M. M. and Poppett, C. and Prada, F. and Rabinowitz, D. and Raichoor, A. and {Ram{\'i}rez-P{\'e}rez}, C. and Rashkovetskyi, M. and Ravoux, C. and Rich, J. and Rocher, A. and Rockosi, C. and Rohlf, J. and {Rom{\'a}n-Herrera}, J. O. and Ross, A. J. and Rossi, G. and Ruggeri, R. and {Ruhlmann-Kleider}, V. and Samushia, L. and Sanchez, E. and Sanders, N. and Schlegel, D. and Schubnell, M. and Seo, H. and Shafieloo, A. and Sharples, R. and Silber, J. and Sinigaglia, F. and Sprayberry, D. and Tan, T. and Tarl{\'e}, G. and Taylor, P. and Turner, W. and {Ure{\~n}a-L{\'o}pez}, L. A. and Vaisakh, R. and Valdes, F. and Valogiannis, G. and {Vargas-Maga{\~n}a}, M. and Verde, L. and Walther, M. and Weaver, B. A. and Weinberg, D. H. and White, M. and Wolfson, M. and Y{\`e}che, C. and Yu, J. and Zaborowski, E. A. and Zarrouk, P. and Zhai, Z. and Zhang, H. and Zhao, C. and Zhao, G. B. and Zhou, R. and Zou, H. and {DESI Collaboration}},
  year = 2025,
  month = oct,
  journal = {Physical Review D},
  volume = {112},
  pages = {083515},
  publisher = {APS},
  issn = {1550-79980556-2821},
  doi = {10.1103/tr6y-kpc6}
}

@article{adameDESI2024VI2025,
  title = {{{DESI}} 2024 {{VI}}: Cosmological Constraints from the Measurements of Baryon Acoustic Oscillations},
  shorttitle = {{{DESI}} 2024 {{VI}}},
  author = {Adame, A. G. and Aguilar, J. and Ahlen, S. and Alam, S. and Alexander, D. M. and Alvarez, M. and Alves, O. and Anand, A. and Andrade, U. and Armengaud, E. and Avila, S. and Aviles, A. and Awan, H. and {Bahr-Kalus}, B. and Bailey, S. and Baltay, C. and Bault, A. and Behera, J. and BenZvi, S. and Bera, A. and Beutler, F. and Bianchi, D. and Blake, C. and Blum, R. and Brieden, S. and Brodzeller, A. and Brooks, D. and {Buckley-Geer}, E. and Burtin, E. and Calderon, R. and Canning, R. and Carnero Rosell, A. and Cereskaite, R. and {Cervantes-Cota}, J. L. and Chabanier, S. and Chaussidon, E. and {Chaves-Montero}, J. and Chen, S. and Chen, X. and Claybaugh, T. and Cole, S. and Cuceu, A. and Davis, T. M. and Dawson, K. and {de la Macorra}, A. and {de Mattia}, A. and Deiosso, N. and Dey, A. and Dey, B. and Ding, Z. and Doel, P. and Edelstein, J. and Eftekharzadeh, S. and Eisenstein, D. J. and Elliott, A. and Fagrelius, P. and Fanning, K. and Ferraro, S. and Ereza, J. and Findlay, N. and Flaugher, B. and {Font-Ribera}, A. and {Forero-S{\'a}nchez}, D. and {Forero-Romero}, J. E. and Frenk, C. S. and {Garcia-Quintero}, C. and Gazta{\~n}aga, E. and {Gil-Mar{\'i}n}, H. and {Gontcho a Gontcho}, S. and {Gonzalez-Morales}, A. X. and {Gonzalez-Perez}, V. and Gordon, C. and Green, D. and Gruen, D. and Gsponer, R. and Gutierrez, G. and Guy, J. and Hadzhiyska, B. and Hahn, C. and Hanif, M. M. S. and {Herrera-Alcantar}, H. K. and Honscheid, K. and Howlett, C. and Huterer, D. and Ir{\v s}i{\v c}, V. and Ishak, M. and Juneau, S. and Kara{\c c}ayl{\i}, N. G. and Kehoe, R. and Kent, S. and Kirkby, D. and Kremin, A. and Krolewski, A. and Lai, Y. and Lan, T.-W. and Landriau, M. and Lang, D. and Lasker, J. and Le Goff, J. M. and Le Guillou, L. and Leauthaud, A. and Levi, M. E. and Li, T. S. and Linder, E. and Lodha, K. and Magneville, C. and Manera, M. and Margala, D. and Martini, P. and Maus, M. and McDonald, P. and {Medina-Varela}, L. and Meisner, A. and {Mena-Fern{\'a}ndez}, J. and Miquel, R. and Moon, J. and Moore, S. and Moustakas, J. and Mueller, E. and {Mu{\~n}oz-Guti{\'e}rrez}, A. and Myers, A. D. and Nadathur, S. and Napolitano, L. and Neveux, R. and Newman, J. A. and Nguyen, N. M. and Nie, J. and Niz, G. and Noriega, H. E. and Padmanabhan, N. and Paillas, E. and {Palanque-Delabrouille}, N. and Pan, J. and Penmetsa, S. and Percival, W. J. and Pieri, M. M. and Pinon, M. and Poppett, C. and Porredon, A. and Prada, F. and {P{\'e}rez-Fern{\'a}ndez}, A. and {P{\'e}rez-R{\`a}fols}, I. and Rabinowitz, D. and Raichoor, A. and {Ram{\'i}rez-P{\'e}rez}, C. and {Ramirez-Solano}, S. and Rashkovetskyi, M. and Ravoux, C. and Rezaie, M. and Rich, J. and Rocher, A. and Rockosi, C. and Roe, N. A. and {Rosado-Marin}, A. and Ross, A. J. and Rossi, G. and Ruggeri, R. and {Ruhlmann-Kleider}, V. and Samushia, L. and Sanchez, E. and Saulder, C. and Schlafly, E. F. and Schlegel, D. and Schubnell, M. and Seo, H. and Shafieloo, A. and Sharples, R. and Silber, J. and Slosar, A. and Smith, A. and Sprayberry, D. and Tan, T. and Tarl{\'e}, G. and Taylor, P. and Trusov, S. and {Ure{\~n}a-L{\'o}pez}, L. A. and Vaisakh, R. and Valcin, D. and Valdes, F. and {Vargas-Maga{\~n}a}, M. and Verde, L. and Walther, M. and Wang, B. and Wang, M. S. and Weaver, B. A. and Weaverdyck, N. and Wechsler, R. H. and Weinberg, D. H. and White, M. and Yu, J. and Yu, Y. and Yuan, S. and Y{\`e}che, C. and Zaborowski, E. A. and Zarrouk, P. and Zhang, H. and Zhao, C. and Zhao, R. and Zhou, R. and Zhuang, T.},
  year = 2025,
  month = feb,
  journal = {Journal of Cosmology and Astroparticle Physics},
  volume = {2025},
  pages = {021},
  publisher = {IOP},
  issn = {1475-7516},
  doi = {10.1088/1475-7516/2025/02/021}
}

@article{amenoucheZTFSNIa2025,
  title = {{{ZTF SN Ia DR2}}: {{Simulations}} and Volume-Limited Sample},
  shorttitle = {{{ZTF SN Ia DR2}}},
  author = {Amenouche, M. and Rosnet, P. and Smith, M. and Rigault, M. and Aubert, M. and {Barjou-Delayre}, C. and Burgaz, U. and Carreres, B. and Dimitriadis, G. and Feinstein, F. and Galbany, L. and Ginolin, M. and Goobar, A. and Harvey, L. and Johansson, J. and Kim, Y.-L. and Maguire, K. and {M{\"u}ller-Bravo}, T. E. and Nordin, J. and Nugent, P. and Racine, B. and Rosselli, D. and Regnault, N. and Sollerman, J. and Terwel, J. H. and Townsend, A. and Groom, S. L. and Kulkarni, S. R. and Kasliwal, M. and Laher, R. R. and Purdum, J.},
  year = 2025,
  month = feb,
  volume = {694},
  pages = {A3},
  publisher = {EDP},
  issn = {0004-6361},
  doi = {10.1051/0004-6361/202452134},
  journal = {\aap}
}

@article{antilogusBrighterfatterEffectPixel2014,
  title = {The Brighter-Fatter Effect and Pixel Correlations in {{CCD}} Sensors},
  author = {Antilogus, P. and Astier, P. and Doherty, P. and Guyonnet, A. and Regnault, N.},
  year = 2014,
  month = mar,
  journal = {Journal of Instrumentation},
  volume = {9},
  pages = {C03048},
  publisher = {IOP},
  doi = {10.1088/1748-0221/9/03/C03048}
}

@article{astierCorrectionBrighterfatterEffect2023,
  title = {Correction of the Brighter-Fatter Effect on the {{CCDs}} of {{Hyper Suprime-Cam}}},
  author = {Astier, Pierre and Regnault, Nicolas},
  year = 2023,
  month = feb,
  volume = {670},
  pages = {A118},
  publisher = {EDP},
  issn = {0004-6361},
  doi = {10.1051/0004-6361/202245407},
  journal = {\aap}
}

@article{astierPhotometrySupernovaeImage2013,
  title = {Photometry of Supernovae in an Image Series: Methods and Application to the {{SuperNova Legacy Survey}} ({{SNLS}})},
  shorttitle = {Photometry of Supernovae in an Image Series},
  author = {Astier, P. and El Hage, P. and Guy, J. and Hardin, D. and Betoule, M. and Fabbro, S. and Fourmanoit, N. and Pain, R. and Regnault, N.},
  year = 2013,
  month = sep,
  volume = {557},
  pages = {A55},
  publisher = {EDP},
  issn = {0004-6361},
  doi = {10.1051/0004-6361/201321668},
  journal = {\aap}
}

@article{astierSupernovaLegacySurvey2006,
  title = {The {{Supernova Legacy Survey}}: Measurement of {{\textohm M}}, {{\textohm$\Lambda$}} and w from the First Year Data Set},
  shorttitle = {The {{Supernova Legacy Survey}}},
  author = {Astier, P. and Guy, J. and Regnault, N. and Pain, R. and Aubourg, E. and Balam, D. and Basa, S. and Carlberg, R. G. and Fabbro, S. and Fouchez, D. and Hook, I. M. and Howell, D. A. and Lafoux, H. and Neill, J. D. and {Palanque-Delabrouille}, N. and Perrett, K. and Pritchet, C. J. and Rich, J. and Sullivan, M. and Taillet, R. and Aldering, G. and Antilogus, P. and Arsenijevic, V. and Balland, C. and Baumont, S. and Bronder, J. and Courtois, H. and Ellis, R. S. and Filiol, M. and Gon{\c c}alves, A. C. and Goobar, A. and Guide, D. and Hardin, D. and Lusset, V. and Lidman, C. and McMahon, R. and Mouchet, M. and Mourao, A. and Perlmutter, S. and Ripoche, P. and Tao, C. and Walton, N.},
  year = 2006,
  month = feb,
  volume = {447},
  pages = {31--48},
  publisher = {EDP},
  issn = {0004-6361},
  doi = {10.1051/0004-6361:20054185},
  journal = {\aap}
}

@article{barbarySNCosmo2024,
  title = {{{SNCosmo}}},
  author = {Barbary, Kyle and Bailey, Stephen and Barentsen, Geert and Barclay, Tom and Biswas, Rahul and Boone, Kyle and Craig, Matt and Feindt, Ulrich and Friesen, Brian and Goldstein, Danny and Jha, Saurabh W. and Jones, David O. and Mondon, Florian and Papadogiannakis, Sem{\'e}li and Perrefort, Daniel and Pierel, Justin and Rodney, Steve and Rose, Benjamin and Saunders, Clare and Sip{\H o}cz, Brigitta and Sofiatti, Caroline and Thomas, Rollin C. and {van Santen}, Jakob and Vincenzi, Maria and Wang, David and {Wood-Vasey}, Michael},
  year = 2024,
  month = feb,
  journal = {Zenodo},
  publisher = {Zenodo},
  doi = {10.5281/zenodo.10684802}
}

@article{bellmZwickyTransientFacility2019,
  title = {The {{Zwicky Transient Facility}}: {{Surveys}} and {{Scheduler}}},
  shorttitle = {The {{Zwicky Transient Facility}}},
  author = {Bellm, Eric C. and Kulkarni, Shrinivas R. and Barlow, Tom and Feindt, Ulrich and Graham, Matthew J. and Goobar, Ariel and Kupfer, Thomas and Ngeow, Chow-Choong and Nugent, Peter and Ofek, Eran and Prince, Thomas A. and Riddle, Reed and Walters, Richard and Ye, Quan-Zhi},
  year = 2019,
  month = jun,
  volume = {131},
  pages = {068003},
  publisher = {IOP},
  issn = {0004-6280},
  doi = {10.1088/1538-3873/ab0c2a},
  journal = {\pasp}
}

@article{bertinSExtractorSoftwareSource1996,
  title = {{{SExtractor}}: {{Software}} for Source Extraction.},
  shorttitle = {{{SExtractor}}},
  author = {Bertin, E. and Arnouts, S.},
  year = 1996,
  month = jun,
  volume = {117},
  pages = {393--404},
  publisher = {EDP},
  issn = {0365-01380004-6361},
  doi = {10.1051/aas:1996164},
  journal = {\aaps}
}

@article{betouleImprovedCosmologicalConstraints2014,
  title = {Improved Cosmological Constraints from a Joint Analysis of the {{SDSS-II}} and {{SNLS}} Supernova Samples},
  author = {Betoule, M. and Kessler, R. and Guy, J. and Mosher, J. and Hardin, D. and Biswas, R. and Astier, P. and {El-Hage}, P. and Konig, M. and Kuhlmann, S. and Marriner, J. and Pain, R. and Regnault, N. and Balland, C. and Bassett, B. A. and Brown, P. J. and Campbell, H. and Carlberg, R. G. and {Cellier-Holzem}, F. and Cinabro, D. and Conley, A. and D'Andrea, C. B. and DePoy, D. L. and Doi, M. and Ellis, R. S. and Fabbro, S. and Filippenko, A. V. and Foley, R. J. and Frieman, J. A. and Fouchez, D. and Galbany, L. and Goobar, A. and Gupta, R. R. and Hill, G. J. and Hlozek, R. and Hogan, C. J. and Hook, I. M. and Howell, D. A. and Jha, S. W. and Le Guillou, L. and Leloudas, G. and Lidman, C. and Marshall, J. L. and M{\"o}ller, A. and Mour{\~a}o, A. M. and Neveu, J. and Nichol, R. and Olmstead, M. D. and {Palanque-Delabrouille}, N. and Perlmutter, S. and Prieto, J. L. and Pritchet, C. J. and Richmond, M. and Riess, A. G. and {Ruhlmann-Kleider}, V. and Sako, M. and Schahmaneche, K. and Schneider, D. P. and Smith, M. and Sollerman, J. and Sullivan, M. and Walton, N. A. and Wheeler, C. J.},
  year = 2014,
  month = aug,
  volume = {568},
  pages = {A22},
  publisher = {EDP},
  issn = {0004-6361},
  doi = {10.1051/0004-6361/201423413},
  journal = {\aap}
}

@article{blagorodnovaSEDMachineRobotic2018,
  title = {The {{SED Machine}}: {{A Robotic Spectrograph}} for {{Fast Transient Classification}}},
  shorttitle = {The {{SED Machine}}},
  author = {Blagorodnova, Nadejda and Neill, James D. and Walters, Richard and Kulkarni, Shrinivas R. and Fremling, Christoffer and {Ben-Ami}, Sagi and Dekany, Richard G. and Fucik, Jason R. and Konidaris, Nick and Nash, Reston and Ngeow, Chow-Choong and Ofek, Eran O. and O' Sullivan, Donal and Quimby, Robert and Ritter, Andreas and Vyhmeister, Karl E.},
  year = 2018,
  month = mar,
  volume = {130},
  pages = {035003},
  publisher = {IOP},
  issn = {0004-6280},
  doi = {10.1088/1538-3873/aaa53f},
  journal = {\pasp}
}

@article{bohlinNewGridsPurehydrogen2020,
  title = {New {{Grids}} of {{Pure-hydrogen White Dwarf NLTE Model Atmospheres}} and the {{HST}}/{{STIS Flux Calibration}}},
  author = {Bohlin, Ralph C. and Hubeny, Ivan and Rauch, Thomas},
  year = 2020,
  month = jul,
  volume = {160},
  pages = {21},
  publisher = {IOP},
  issn = {0004-6256},
  doi = {10.3847/1538-3881/ab94b4},
  journal = {\aj}
}

@article{bohlinTechniquesReviewAbsolute2014,
  title = {Techniques and {{Review}} of {{Absolute Flux Calibration}} from the {{Ultraviolet}} to the {{Mid-Infrared}}},
  author = {Bohlin, Ralph C. and Gordon, Karl D. and Tremblay, P.-E.},
  year = 2014,
  month = aug,
  volume = {126},
  pages = {711},
  publisher = {IOP},
  issn = {0004-6280},
  doi = {10.1086/677655},
  journal = {\pasp}
}

@article{broutPantheonAnalysisCosmological2022,
  title = {The {{Pantheon}}+ {{Analysis}}: {{Cosmological Constraints}}},
  shorttitle = {The {{Pantheon}}+ {{Analysis}}},
  author = {Brout, Dillon and Scolnic, Dan and Popovic, Brodie and Riess, Adam G. and Carr, Anthony and Zuntz, Joe and Kessler, Rick and Davis, Tamara M. and Hinton, Samuel and Jones, David and Kenworthy, W. D'Arcy and Peterson, Erik R. and Said, Khaled and Taylor, Georgie and Ali, Noor and Armstrong, Patrick and Charvu, Pranav and Dwomoh, Arianna and Meldorf, Cole and Palmese, Antonella and Qu, Helen and Rose, Benjamin M. and Sanchez, Bruno and Stubbs, Christopher W. and Vincenzi, Maria and Wood, Charlotte M. and Brown, Peter J. and Chen, Rebecca and Chambers, Ken and Coulter, David A. and Dai, Mi and Dimitriadis, Georgios and Filippenko, Alexei V. and Foley, Ryan J. and Jha, Saurabh W. and Kelsey, Lisa and Kirshner, Robert P. and M{\"o}ller, Anais and Muir, Jessie and Nadathur, Seshadri and Pan, Yen-Chen and Rest, Armin and {Rojas-Bravo}, Cesar and Sako, Masao and Siebert, Matthew R. and Smith, Mat and Stahl, Benjamin E. and Wiseman, Phil},
  year = 2022,
  month = oct,
  volume = {938},
  pages = {110},
  publisher = {IOP},
  issn = {0004-637X},
  doi = {10.3847/1538-4357/ac8e04},
  journal = {\apj}
}

@book{daskdevelopmentteamDaskLibraryDynamic2016,
  title = {Dask: {{Library}} for Dynamic Task Scheduling},
  author = {{Dask Development Team}},
  year = 2016
}

@article{dekanyZwickyTransientFacility2020,
  title = {The {{Zwicky Transient Facility}}: {{Observing System}}},
  shorttitle = {The {{Zwicky Transient Facility}}},
  author = {Dekany, Richard and Smith, Roger M. and Riddle, Reed and Feeney, Michael and Porter, Michael and Hale, David and Zolkower, Jeffry and Belicki, Justin and Kaye, Stephen and Henning, John and Walters, Richard and Cromer, John and Delacroix, Alex and Rodriguez, Hector and Reiley, Daniel J. and Mao, Peter and Hover, David and Murphy, Patrick and Burruss, Rick and Baker, John and Kowalski, Marek and Reif, Klaus and Mueller, Phillip and Bellm, Eric and Graham, Matthew and Kulkarni, Shrinivas R.},
  year = 2020,
  month = mar,
  volume = {132},
  pages = {038001},
  publisher = {IOP},
  issn = {0004-6280},
  doi = {10.1088/1538-3873/ab4ca2},
  journal = {\pasp}
}

@article{descollaborationDarkEnergySurvey2024a,
  title = {The {{Dark Energy Survey}}: {{Cosmology Results}} with {$\sim$}1500 {{New High-redshift Type Ia Supernovae Using}} the {{Full}} 5 Yr {{Data Set}}},
  shorttitle = {The {{Dark Energy Survey}}},
  author = {{DES Collaboration} and Abbott, T. M. C. and Acevedo, M. and Aguena, M. and Alarcon, A. and Allam, S. and Alves, O. and Amon, A. and {Andrade-Oliveira}, F. and Annis, J. and Armstrong, P. and Asorey, J. and Avila, S. and Bacon, D. and Bassett, B. A. and Bechtol, K. and Bernardinelli, P. H. and Bernstein, G. M. and Bertin, E. and Blazek, J. and Bocquet, S. and Brooks, D. and Brout, D. and {Buckley-Geer}, E. and Burke, D. L. and Camacho, H. and Camilleri, R. and Campos, A. and Carnero Rosell, A. and Carollo, D. and Carr, A. and Carretero, J. and Castander, F. J. and Cawthon, R. and Chang, C. and Chen, R. and Choi, A. and Conselice, C. and Costanzi, M. and {da Costa}, L. N. and Crocce, M. and Davis, T. M. and DePoy, D. L. and Desai, S. and Diehl, H. T. and Dixon, M. and Dodelson, S. and Doel, P. and Doux, C. and {Drlica-Wagner}, A. and {Elvin-Poole}, J. and Everett, S. and Ferrero, I. and Fert{\'e}, A. and Flaugher, B. and Foley, R. J. and Fosalba, P. and Friedel, D. and Frieman, J. and Frohmaier, C. and Galbany, L. and {Garc{\'i}a-Bellido}, J. and Gatti, M. and Gaztanaga, E. and Giannini, G. and Glazebrook, K. and Graur, O. and Gruen, D. and Gruendl, R. A. and Gutierrez, G. and Hartley, W. G. and Herner, K. and Hinton, S. R. and Hollowood, D. L. and Honscheid, K. and Huterer, D. and Jain, B. and James, D. J. and Jeffrey, N. and Kasai, E. and Kelsey, L. and Kent, S. and Kessler, R. and Kim, A. G. and Kirshner, R. P. and Kovacs, E. and Kuehn, K. and Lahav, O. and Lee, J. and Lee, S. and Lewis, G. F. and Li, T. S. and Lidman, C. and Lin, H. and Malik, U. and Marshall, J. L. and Martini, P. and {Mena-Fern{\'a}ndez}, J. and Menanteau, F. and Miquel, R. and Mohr, J. J. and Mould, J. and Muir, J. and M{\"o}ller, A. and Neilsen, E. and Nichol, R. C. and Nugent, P. and Ogando, R. L. C. and Palmese, A. and Pan, Y.-C. and Paterno, M. and Percival, W. J. and Pereira, M. E. S. and Pieres, A. and Malag{\'o}n, A. A. Plazas and Popovic, B. and Porredon, A. and Prat, J. and Qu, H. and Raveri, M. and {Rodr{\'i}guez-Monroy}, M. and Romer, A. K. and Roodman, A. and Rose, B. and Sako, M. and Sanchez, E. and Sanchez Cid, D. and Schubnell, M. and Scolnic, D. and {Sevilla-Noarbe}, I. and Shah, P. and Smith, J. {\relax Allyn}. and Smith, M. and {Soares-Santos}, M. and Suchyta, E. and Sullivan, M. and Suntzeff, N. and Swanson, M. E. C. and S{\'a}nchez, B. O. and Tarle, G. and Taylor, G. and Thomas, D. and To, C. and Toy, M. and Troxel, M. A. and Tucker, B. E. and Tucker, D. L. and Uddin, S. A. and Vincenzi, M. and Walker, A. R. and Weaverdyck, N. and Wechsler, R. H. and Weller, J. and Wester, W. and Wiseman, P. and Yamamoto, M. and Yuan, F. and Zhang, B. and Zhang, Y.},
  year = 2024,
  month = sep,
  volume = {973},
  pages = {L14},
  publisher = {IOP},
  issn = {0004-637X},
  doi = {10.3847/2041-8213/ad6f9f},
  journal = {\apj}
}

@article{gaiacollaborationGaiaEarlyData2021,
  title = {Gaia {{Early Data Release}} 3. {{Summary}} of the Contents and Survey Properties},
  author = {{Gaia Collaboration} and Brown, A. G. A. and Vallenari, A. and Prusti, T. and {de Bruijne}, J. H. J. and Babusiaux, C. and Biermann, M. and Creevey, O. L. and Evans, D. W. and Eyer, L. and Hutton, A. and Jansen, F. and Jordi, C. and Klioner, S. A. and Lammers, U. and Lindegren, L. and Luri, X. and Mignard, F. and Panem, C. and Pourbaix, D. and Randich, S. and Sartoretti, P. and Soubiran, C. and Walton, N. A. and Arenou, F. and {Bailer-Jones}, C. A. L. and Bastian, U. and Cropper, M. and Drimmel, R. and Katz, D. and Lattanzi, M. G. and {van Leeuwen}, F. and Bakker, J. and Cacciari, C. and Casta{\~n}eda, J. and De Angeli, F. and Ducourant, C. and Fabricius, C. and Fouesneau, M. and Fr{\'e}mat, Y. and Guerra, R. and Guerrier, A. and Guiraud, J. and {Jean-Antoine Piccolo}, A. and Masana, E. and Messineo, R. and Mowlavi, N. and Nicolas, C. and Nienartowicz, K. and Pailler, F. and Panuzzo, P. and Riclet, F. and Roux, W. and Seabroke, G. M. and Sordo, R. and Tanga, P. and Th{\'e}venin, F. and {Gracia-Abril}, G. and Portell, J. and Teyssier, D. and Altmann, M. and Andrae, R. and {Bellas-Velidis}, I. and Benson, K. and Berthier, J. and Blomme, R. and Brugaletta, E. and Burgess, P. W. and Busso, G. and Carry, B. and Cellino, A. and Cheek, N. and Clementini, G. and Damerdji, Y. and Davidson, M. and Delchambre, L. and Dell'Oro, A. and {Fern{\'a}ndez-Hern{\'a}ndez}, J. and Galluccio, L. and {Garc{\'i}a-Lario}, P. and {Garcia-Reinaldos}, M. and {Gonz{\'a}lez-N{\'u}{\~n}ez}, J. and Gosset, E. and Haigron, R. and Halbwachs, J.-L. and Hambly, N. C. and Harrison, D. L. and Hatzidimitriou, D. and Heiter, U. and Hern{\'a}ndez, J. and Hestroffer, D. and Hodgkin, S. T. and Holl, B. and Jan{\ss}en, K. and {Jevardat de Fombelle}, G. and Jordan, S. and {Krone-Martins}, A. and Lanzafame, A. C. and L{\"o}ffler, W. and Lorca, A. and Manteiga, M. and Marchal, O. and Marrese, P. M. and Moitinho, A. and Mora, A. and Muinonen, K. and Osborne, P. and Pancino, E. and Pauwels, T. and Petit, J.-M. and {Recio-Blanco}, A. and Richards, P. J. and Riello, M. and Rimoldini, L. and Robin, A. C. and Roegiers, T. and Rybizki, J. and Sarro, L. M. and Siopis, C. and Smith, M. and Sozzetti, A. and Ulla, A. and Utrilla, E. and {van Leeuwen}, M. and {van Reeven}, W. and Abbas, U. and Abreu Aramburu, A. and Accart, S. and Aerts, C. and Aguado, J. J. and Ajaj, M. and Altavilla, G. and {\'A}lvarez, M. A. and {\'A}lvarez Cid-Fuentes, J. and Alves, J. and Anderson, R. I. and Anglada Varela, E. and Antoja, T. and Audard, M. and Baines, D. and Baker, S. G. and {Balaguer-N{\'u}{\~n}ez}, L. and Balbinot, E. and Balog, Z. and Barache, C. and Barbato, D. and Barros, M. and Barstow, M. A. and Bartolom{\'e}, S. and Bassilana, J.-L. and Bauchet, N. and {Baudesson-Stella}, A. and Becciani, U. and Bellazzini, M. and Bernet, M. and Bertone, S. and Bianchi, L. and {Blanco-Cuaresma}, S. and Boch, T. and Bombrun, A. and Bossini, D. and Bouquillon, S. and Bragaglia, A. and Bramante, L. and Breedt, E. and Bressan, A. and Brouillet, N. and Bucciarelli, B. and Burlacu, A. and Busonero, D. and Butkevich, A. G. and Buzzi, R. and Caffau, E. and Cancelliere, R. and C{\'a}novas, H. and {Cantat-Gaudin}, T. and Carballo, R. and Carlucci, T. and Carnerero, M. I. and Carrasco, J. M. and Casamiquela, L. and Castellani, M. and {Castro-Ginard}, A. and Castro Sampol, P. and Chaoul, L. and Charlot, P. and Chemin, L. and Chiavassa, A. and Cioni, M.-R. L. and Comoretto, G. and Cooper, W. J. and Cornez, T. and Cowell, S. and Crifo, F. and Crosta, M. and Crowley, C. and Dafonte, C. and Dapergolas, A. and David, M. and David, P.},
  year = 2021,
  month = may,
  volume = {649},
  pages = {A1},
  publisher = {EDP},
  issn = {0004-6361},
  doi = {10.1051/0004-6361/202039657},
  journal = {\aap}
}

@article{gaiacollaborationGaiaMission2016,
  title = {The {{Gaia}} Mission},
  author = {{Gaia Collaboration} and Prusti, T. and {de Bruijne}, J. H. J. and Brown, A. G. A. and Vallenari, A. and Babusiaux, C. and {Bailer-Jones}, C. A. L. and Bastian, U. and Biermann, M. and Evans, D. W. and Eyer, L. and Jansen, F. and Jordi, C. and Klioner, S. A. and Lammers, U. and Lindegren, L. and Luri, X. and Mignard, F. and Milligan, D. J. and Panem, C. and Poinsignon, V. and Pourbaix, D. and Randich, S. and Sarri, G. and Sartoretti, P. and Siddiqui, H. I. and Soubiran, C. and Valette, V. and {van Leeuwen}, F. and Walton, N. A. and Aerts, C. and Arenou, F. and Cropper, M. and Drimmel, R. and H{\o}g, E. and Katz, D. and Lattanzi, M. G. and O'Mullane, W. and Grebel, E. K. and Holland, A. D. and Huc, C. and Passot, X. and Bramante, L. and Cacciari, C. and Casta{\~n}eda, J. and Chaoul, L. and Cheek, N. and De Angeli, F. and Fabricius, C. and Guerra, R. and Hern{\'a}ndez, J. and {Jean-Antoine-Piccolo}, A. and Masana, E. and Messineo, R. and Mowlavi, N. and Nienartowicz, K. and {Ord{\'o}{\~n}ez-Blanco}, D. and Panuzzo, P. and Portell, J. and Richards, P. J. and Riello, M. and Seabroke, G. M. and Tanga, P. and Th{\'e}venin, F. and Torra, J. and Els, S. G. and {Gracia-Abril}, G. and Comoretto, G. and {Garcia-Reinaldos}, M. and Lock, T. and Mercier, E. and Altmann, M. and Andrae, R. and Astraatmadja, T. L. and {Bellas-Velidis}, I. and Benson, K. and Berthier, J. and Blomme, R. and Busso, G. and Carry, B. and Cellino, A. and Clementini, G. and Cowell, S. and Creevey, O. and Cuypers, J. and Davidson, M. and De Ridder, J. and {de Torres}, A. and Delchambre, L. and Dell'Oro, A. and Ducourant, C. and Fr{\'e}mat, Y. and {Garc{\'i}a-Torres}, M. and Gosset, E. and Halbwachs, J.-L. and Hambly, N. C. and Harrison, D. L. and Hauser, M. and Hestroffer, D. and Hodgkin, S. T. and Huckle, H. E. and Hutton, A. and Jasniewicz, G. and Jordan, S. and Kontizas, M. and Korn, A. J. and Lanzafame, A. C. and Manteiga, M. and Moitinho, A. and Muinonen, K. and Osinde, J. and Pancino, E. and Pauwels, T. and Petit, J.-M. and {Recio-Blanco}, A. and Robin, A. C. and Sarro, L. M. and Siopis, C. and Smith, M. and Smith, K. W. and Sozzetti, A. and Thuillot, W. and {van Reeven}, W. and Viala, Y. and Abbas, U. and Abreu Aramburu, A. and Accart, S. and Aguado, J. J. and Allan, P. M. and Allasia, W. and Altavilla, G. and {\'A}lvarez, M. A. and Alves, J. and Anderson, R. I. and Andrei, A. H. and Anglada Varela, E. and Antiche, E. and Antoja, T. and Ant{\'o}n, S. and Arcay, B. and Atzei, A. and Ayache, L. and Bach, N. and Baker, S. G. and {Balaguer-N{\'u}{\~n}ez}, L. and Barache, C. and Barata, C. and Barbier, A. and Barblan, F. and Baroni, M. and {Barrado y Navascu{\'e}s}, D. and Barros, M. and Barstow, M. A. and Becciani, U. and Bellazzini, M. and Bellei, G. and Bello Garc{\'i}a, A. and Belokurov, V. and Bendjoya, P. and Berihuete, A. and Bianchi, L. and Bienaym{\'e}, O. and Billebaud, F. and Blagorodnova, N. and {Blanco-Cuaresma}, S. and Boch, T. and Bombrun, A. and Borrachero, R. and Bouquillon, S. and Bourda, G. and Bouy, H. and Bragaglia, A. and Breddels, M. A. and Brouillet, N. and Br{\"u}semeister, T. and Bucciarelli, B. and Budnik, F. and Burgess, P. and Burgon, R. and Burlacu, A. and Busonero, D. and Buzzi, R. and Caffau, E. and Cambras, J. and Campbell, H. and Cancelliere, R. and {Cantat-Gaudin}, T. and Carlucci, T. and Carrasco, J. M. and Castellani, M. and Charlot, P. and Charnas, J. and Charvet, P. and Chassat, F. and Chiavassa, A. and Clotet, M. and Cocozza, G. and Collins, R. S. and Collins, P. and Costigan, G.},
  year = 2016,
  month = nov,
  volume = {595},
  pages = {A1},
  publisher = {EDP},
  issn = {0004-6361},
  doi = {10.1051/0004-6361/201629272},
  journal = {\aap}
}

@article{grahamZwickyTransientFacility2019,
  title = {The {{Zwicky Transient Facility}}: {{Science Objectives}}},
  shorttitle = {The {{Zwicky Transient Facility}}},
  author = {Graham, Matthew J. and Kulkarni, S. R. and Bellm, Eric C. and Adams, Scott M. and Barbarino, Cristina and Blagorodnova, Nadejda and Bodewits, Dennis and Bolin, Bryce and Brady, Patrick R. and Cenko, S. Bradley and Chang, Chan-Kao and Coughlin, Michael W. and De, Kishalay and Eadie, Gwendolyn and Farnham, Tony L. and Feindt, Ulrich and Franckowiak, Anna and Fremling, Christoffer and Gezari, Suvi and Ghosh, Shaon and Goldstein, Daniel A. and Golkhou, V. Zach and Goobar, Ariel and Ho, Anna Y. Q. and Huppenkothen, Daniela and Ivezi{\'c}, {\v Z}eljko and Jones, R. Lynne and Juric, Mario and Kaplan, David L. and Kasliwal, Mansi M. and Kelley, Michael S. P. and Kupfer, Thomas and Lee, Chien-De and Lin, Hsing Wen and Lunnan, Ragnhild and Mahabal, Ashish A. and Miller, Adam A. and Ngeow, Chow-Choong and Nugent, Peter and Ofek, Eran O. and Prince, Thomas A. and Rauch, Ludwig and {van Roestel}, Jan and Schulze, Steve and Singer, Leo P. and Sollerman, Jesper and Taddia, Francesco and Yan, Lin and Ye, Quan-Zhi and Yu, Po-Chieh and Barlow, Tom and Bauer, James and Beck, Ron and Belicki, Justin and Biswas, Rahul and Brinnel, Valery and Brooke, Tim and Bue, Brian and Bulla, Mattia and Burruss, Rick and Connolly, Andrew and Cromer, John and Cunningham, Virginia and Dekany, Richard and Delacroix, Alex and Desai, Vandana and Duev, Dmitry A. and Feeney, Michael and Flynn, David and Frederick, Sara and {Gal-Yam}, Avishay and Giomi, Matteo and Groom, Steven and Hacopians, Eugean and Hale, David and Helou, George and Henning, John and Hover, David and Hillenbrand, Lynne A. and Howell, Justin and Hung, Tiara and Imel, David and Ip, Wing-Huen and Jackson, Edward and Kaspi, Shai and Kaye, Stephen and Kowalski, Marek and Kramer, Emily and Kuhn, Michael and Landry, Walter and Laher, Russ R. and Mao, Peter and Masci, Frank J. and Monkewitz, Serge and Murphy, Patrick and Nordin, Jakob and Patterson, Maria T. and Penprase, Bryan and Porter, Michael and Rebbapragada, Umaa and Reiley, Dan and Riddle, Reed and Rigault, Mickael and Rodriguez, Hector and Rusholme, Ben and {van Santen}, Jakob and Shupe, David L. and Smith, Roger M. and Soumagnac, Maayane T. and Stein, Robert and Surace, Jason and Szkody, Paula and Terek, Scott and Van Sistine, Angela and {van Velzen}, Sjoert and Vestrand, W. Thomas and Walters, Richard and Ward, Charlotte and Zhang, Chaoran and Zolkower, Jeffry},
  year = 2019,
  month = jul,
  volume = {131},
  pages = {078001},
  publisher = {IOP},
  issn = {0004-6280},
  doi = {10.1088/1538-3873/ab006c},
  journal = {\pasp}
}

@article{guyonnetEvidenceSelfinteractionCharge2015,
  title = {Evidence for Self-Interaction of Charge Distribution in Charge-Coupled Devices},
  author = {Guyonnet, A. and Astier, P. and Antilogus, P. and Regnault, N. and Doherty, P.},
  year = 2015,
  month = mar,
  volume = {575},
  pages = {A41},
  publisher = {EDP},
  issn = {0004-6361},
  doi = {10.1051/0004-6361/201424897},
  journal = {\aap}
}

@article{guySALT2UsingDistant2007,
  title = {{{SALT2}}: Using Distant Supernovae to Improve the Use of Type {{Ia}} Supernovae as Distance Indicators},
  shorttitle = {{{SALT2}}},
  author = {Guy, J. and Astier, P. and Baumont, S. and Hardin, D. and Pain, R. and Regnault, N. and Basa, S. and Carlberg, R. G. and Conley, A. and Fabbro, S. and Fouchez, D. and Hook, I. M. and Howell, D. A. and Perrett, K. and Pritchet, C. J. and Rich, J. and Sullivan, M. and Antilogus, P. and Aubourg, E. and Bazin, G. and Bronder, J. and Filiol, M. and {Palanque-Delabrouille}, N. and Ripoche, P. and {Ruhlmann-Kleider}, V.},
  year = 2007,
  month = apr,
  volume = {466},
  pages = {11--21},
  publisher = {EDP},
  issn = {0004-6361},
  doi = {10.1051/0004-6361:20066930},
  journal = {\aap}
}

@article{guySupernovaLegacySurvey2010,
  title = {The {{Supernova Legacy Survey}} 3-Year Sample: {{Type Ia}} Supernovae Photometric Distances and Cosmological Constraints},
  shorttitle = {The {{Supernova Legacy Survey}} 3-Year Sample},
  author = {Guy, J. and Sullivan, M. and Conley, A. and Regnault, N. and Astier, P. and Balland, C. and Basa, S. and Carlberg, R. G. and Fouchez, D. and Hardin, D. and Hook, I. M. and Howell, D. A. and Pain, R. and {Palanque-Delabrouille}, N. and Perrett, K. M. and Pritchet, C. J. and Rich, J. and {Ruhlmann-Kleider}, V. and Balam, D. and Baumont, S. and Ellis, R. S. and Fabbro, S. and Fakhouri, H. K. and Fourmanoit, N. and {Gonz{\'a}lez-Gait{\'a}n}, S. and Graham, M. L. and Hsiao, E. and Kronborg, T. and Lidman, C. and Mourao, A. M. and Perlmutter, S. and Ripoche, P. and Suzuki, N. and Walker, E. S.},
  year = 2010,
  month = nov,
  volume = {523},
  pages = {A7},
  publisher = {EDP},
  issn = {0004-6361},
  doi = {10.1051/0004-6361/201014468},
  journal = {\aap}
}

@article{holtzmanSloanDigitalSky2008,
  title = {The {{Sloan Digital Sky Survey-II}}: {{Photometry}} and {{Supernova IA Light Curves}} from the 2005 {{Data}}},
  shorttitle = {The {{Sloan Digital Sky Survey-II}}},
  author = {Holtzman, Jon A. and Marriner, John and Kessler, Richard and Sako, Masao and Dilday, Ben and Frieman, Joshua A. and Schneider, Donald P. and Bassett, Bruce and Becker, Andrew and Cinabro, David and DeJongh, Fritz and Depoy, Darren L. and Doi, Mamoru and Garnavich, Peter M. and Hogan, Craig J. and Jha, Saurabh and Konishi, Kohki and Lampeitl, Hubert and Marshall, Jennifer L. and McGinnis, David and Miknaitis, Gajus and Nichol, Robert C. and Prieto, Jose Luis and Riess, Adam G. and Richmond, Michael W. and Romani, Roger and Smith, Mathew and Takanashi, Naohiro and Tokita, Kouichi and {van der Heyden}, Kurt and Yasuda, Naoki and Zheng, Chen},
  year = 2008,
  month = dec,
  volume = {136},
  pages = {2306--2320},
  publisher = {IOP},
  issn = {0004-6256},
  doi = {10.1088/0004-6256/136/6/2306},
  journal = {\aj}
}

@article{kowalskiImprovedCosmologicalConstraints2008,
  title = {Improved {{Cosmological Constraints}} from {{New}}, {{Old}}, and {{Combined Supernova Data Sets}}},
  author = {Kowalski, M. and Rubin, D. and Aldering, G. and Agostinho, R. J. and Amadon, A. and Amanullah, R. and Balland, C. and Barbary, K. and Blanc, G. and Challis, P. J. and Conley, A. and Connolly, N. V. and Covarrubias, R. and Dawson, K. S. and Deustua, S. E. and Ellis, R. and Fabbro, S. and Fadeyev, V. and Fan, X. and Farris, B. and Folatelli, G. and Frye, B. L. and Garavini, G. and Gates, E. L. and Germany, L. and Goldhaber, G. and Goldman, B. and Goobar, A. and Groom, D. E. and Haissinski, J. and Hardin, D. and Hook, I. and Kent, S. and Kim, A. G. and Knop, R. A. and Lidman, C. and Linder, E. V. and Mendez, J. and Meyers, J. and Miller, G. J. and Moniez, M. and Mour{\~a}o, A. M. and Newberg, H. and Nobili, S. and Nugent, P. E. and Pain, R. and Perdereau, O. and Perlmutter, S. and Phillips, M. M. and Prasad, V. and Quimby, R. and Regnault, N. and Rich, J. and Rubenstein, E. P. and {Ruiz-Lapuente}, P. and Santos, F. D. and Schaefer, B. E. and Schommer, R. A. and Smith, R. C. and Soderberg, A. M. and Spadafora, A. L. and Strolger, L.-G. and Strovink, M. and Suntzeff, N. B. and Suzuki, N. and Thomas, R. C. and Walton, N. A. and Wang, L. and {Wood-Vasey}, W. M. and Yun, J. L.},
  year = 2008,
  month = oct,
  volume = {686},
  pages = {749--778},
  publisher = {IOP},
  issn = {0004-637X},
  doi = {10.1086/589937},
  journal = {\apj}
}

@article{lezmyHyperGalHyperspectralScene2022,
  title = {{{HyperGal}}: {{Hyperspectral}} Scene Modeling for Supernova Typing with the {{SED Machine}} Integral Field Spectrograph},
  shorttitle = {{{HyperGal}}},
  author = {Lezmy, J. and Copin, Y. and Rigault, M. and Smith, M. and Neill, J. D.},
  year = 2022,
  month = dec,
  volume = {668},
  pages = {A43},
  publisher = {EDP},
  issn = {0004-6361},
  doi = {10.1051/0004-6361/202244740},
  journal = {\aap}
}

@misc{lsstsciencecollaborationLSSTScienceBook2009,
  title = {{{LSST Science Book}}, {{Version}} 2.0},
  author = {{LSST Science Collaboration} and Abell, Paul A. and Allison, Julius and Anderson, Scott F. and Andrew, John R. and Angel, J. Roger P. and Armus, Lee and Arnett, David and Asztalos, S. J. and Axelrod, Tim S. and Bailey, Stephen and Ballantyne, D. R. and Bankert, Justin R. and Barkhouse, Wayne A. and Barr, Jeffrey D. and Barrientos, L. Felipe and Barth, Aaron J. and Bartlett, James G. and Becker, Andrew C. and Becla, Jacek and Beers, Timothy C. and Bernstein, Joseph P. and Biswas, Rahul and Blanton, Michael R. and Bloom, Joshua S. and Bochanski, John J. and Boeshaar, Pat and Borne, Kirk D. and Bradac, Marusa and Brandt, W. N. and Bridge, Carrie R. and Brown, Michael E. and Brunner, Robert J. and Bullock, James S. and Burgasser, Adam J. and Burge, James H. and Burke, David L. and Cargile, Phillip A. and Chandrasekharan, Srinivasan and Chartas, George and Chesley, Steven R. and Chu, You-Hua and Cinabro, David and Claire, Mark W. and Claver, Charles F. and Clowe, Douglas and Connolly, A. J. and Cook, Kem H. and Cooke, Jeff and Cooray, Asantha and Covey, Kevin R. and Culliton, Christopher S. and {de Jong}, Roelof and {de Vries}, Willem H. and Debattista, Victor P. and Delgado, Francisco and Dell'Antonio, Ian P. and Dhital, Saurav and Di Stefano, Rosanne and Dickinson, Mark and Dilday, Benjamin and Djorgovski, S. G. and Dobler, Gregory and Donalek, Ciro and {Dubois-Felsmann}, Gregory and Durech, Josef and Eliasdottir, Ardis and Eracleous, Michael and Eyer, Laurent and Falco, Emilio E. and Fan, Xiaohui and Fassnacht, Christopher D. and Ferguson, Harry C. and Fernandez, Yanga R. and Fields, Brian D. and Finkbeiner, Douglas and Figueroa, Eduardo E. and Fox, Derek B. and Francke, Harold and Frank, James S. and Frieman, Josh and Fromenteau, Sebastien and Furqan, Muhammad and Galaz, Gaspar and {Gal-Yam}, A. and Garnavich, Peter and Gawiser, Eric and Geary, John and Gee, Perry and Gibson, Robert R. and Gilmore, Kirk and Grace, Emily A. and Green, Richard F. and Gressler, William J. and Grillmair, Carl J. and Habib, Salman and Haggerty, J. S. and Hamuy, Mario and Harris, Alan W. and Hawley, Suzanne L. and Heavens, Alan F. and Hebb, Leslie and Henry, Todd J. and Hileman, Edward and Hilton, Eric J. and Hoadley, Keri and Holberg, J. B. and Holman, Matt J. and Howell, Steve B. and Infante, Leopoldo and Ivezic, Zeljko and Jacoby, Suzanne H. and Jain, Bhuvnesh and {R} and {Jedicke} and Jee, M. James and Garrett Jernigan, J. and Jha, Saurabh W. and Johnston, Kathryn V. and Jones, R. Lynne and Juric, Mario and Kaasalainen, Mikko and {Styliani} and {Kafka} and Kahn, Steven M. and Kaib, Nathan A. and Kalirai, Jason and Kantor, Jeff and Kasliwal, Mansi M. and Keeton, Charles R. and Kessler, Richard and Knezevic, Zoran and Kowalski, Adam and Krabbendam, Victor L. and Krughoff, K. Simon and Kulkarni, Shrinivas and Kuhlman, Stephen and Lacy, Mark and Lepine, Sebastien and Liang, Ming and Lien, Amy and Lira, Paulina and Long, Knox S. and Lorenz, Suzanne and Lotz, Jennifer M. and Lupton, R. H. and Lutz, Julie and Macri, Lucas M. and Mahabal, Ashish A. and Mandelbaum, Rachel and Marshall, Phil and May, Morgan and McGehee, Peregrine M. and Meadows, Brian T. and Meert, Alan and Milani, Andrea and Miller, Christopher J. and Miller, Michelle and Mills, David and Minniti, Dante and Monet, David and Mukadam, Anjum S. and Nakar, Ehud and Neill, Douglas R. and Newman, Jeffrey A. and Nikolaev, Sergei and Nordby, Martin and O'Connor, Paul and Oguri, Masamune and Oliver, John and Olivier, Scot S. and Olsen, Julia K. and Olsen, Knut and Olszewski, Edward W. and Oluseyi, Hakeem and Padilla, Nelson D. and Parker, Alex and Pepper, Joshua and Peterson, John R. and Petry, Catherine and Pinto, Philip A. and Pizagno, James L. and Popescu, Bogdan and Prsa, Andrej and Radcka, Veljko and Raddick, M. Jordan and Rasmussen, Andrew and Rau, Arne and Rho, Jeonghee and Rhoads, James E. and Richards, Gordon T. and Ridgway, Stephen T. and Robertson, Brant E. and Roskar, Rok and Saha, Abhijit and Sarajedini, Ata and Scannapieco, Evan and Schalk, Terry and Schindler, Rafe and Schmidt, Samuel},
  year = 2009,
  month = dec,
  publisher = {arXiv},
  doi = {10.48550/arXiv.0912.0201}
}

@article{magnierPanSTARRSPhotometricAstrometric2020,
  title = {Pan-{{STARRS Photometric}} and {{Astrometric Calibration}}},
  author = {Magnier, {\relax Eugene}. A. and Schlafly, {\relax Edward}. F. and Finkbeiner, Douglas P. and Tonry, J. L. and Goldman, B. and R{\"o}ser, S. and Schilbach, E. and Casertano, S. and Chambers, K. C. and Flewelling, H. A. and Huber, M. E. and Price, P. A. and Sweeney, W. E. and Waters, C. Z. and Denneau, L. and Draper, P. W. and Hodapp, K. W. and Jedicke, R. and Kaiser, N. and Kudritzki, R.-P. and Metcalfe, N. and Stubbs, C. W. and Wainscoat, R. J.},
  year = 2020,
  month = nov,
  volume = {251},
  pages = {6},
  publisher = {IOP},
  issn = {0067-0049},
  doi = {10.3847/1538-4365/abb82a},
  journal = {\apjs}
}

@article{masciZwickyTransientFacility2019,
  title = {The {{Zwicky Transient Facility}}: {{Data Processing}}, {{Products}}, and {{Archive}}},
  shorttitle = {The {{Zwicky Transient Facility}}},
  author = {Masci, Frank J. and Laher, Russ R. and Rusholme, Ben and Shupe, David L. and Groom, Steven and Surace, Jason and Jackson, Edward and Monkewitz, Serge and Beck, Ron and Flynn, David and Terek, Scott and Landry, Walter and Hacopians, Eugean and Desai, Vandana and Howell, Justin and Brooke, Tim and Imel, David and Wachter, Stefanie and Ye, Quan-Zhi and Lin, Hsing-Wen and Cenko, S. Bradley and Cunningham, Virginia and Rebbapragada, Umaa and Bue, Brian and Miller, Adam A. and Mahabal, Ashish and Bellm, Eric C. and Patterson, Maria T. and Juri{\'c}, Mario and Golkhou, V. Zach and Ofek, Eran O. and Walters, Richard and Graham, Matthew and Kasliwal, Mansi M. and Dekany, Richard G. and Kupfer, Thomas and Burdge, Kevin and Cannella, Christopher B. and Barlow, Tom and Van Sistine, Angela and Giomi, Matteo and Fremling, Christoffer and Blagorodnova, Nadejda and Levitan, David and Riddle, Reed and Smith, Roger M. and Helou, George and Prince, Thomas A. and Kulkarni, Shrinivas R.},
  year = 2019,
  month = jan,
  volume = {131},
  pages = {018003},
  publisher = {IOP},
  issn = {0004-6280},
  doi = {10.1088/1538-3873/aae8ac},
  journal = {\pasp}
}

@misc{neveuAccurateMeasurementTelescope2024,
  title = {Accurate Measurement of Telescope Filter Bandpasses with a {{Collimated Beam Projector}} and Impact on Cosmological Parameters},
  author = {Neveu, J{\'e}r{\'e}my and Kuhn, Dylan and Souverin, Thierry and {LEMAITRE collaboration}},
  year = 2024,
  month = jul,
  publisher = {arXiv},
  doi = {10.48550/arXiv.2407.01650}
}

@article{pattersonZwickyTransientFacility2019,
  title = {The {{Zwicky Transient Facility Alert Distribution System}}},
  author = {Patterson, Maria T. and Bellm, Eric C. and Rusholme, Ben and Masci, Frank J. and Juric, Mario and Krughoff, K. Simon and Golkhou, V. Zach and Graham, Matthew J. and Kulkarni, Shrinivas R. and Helou, George and {Zwicky Transient Facility Collaboration}},
  year = 2019,
  month = jan,
  volume = {131},
  pages = {018001},
  publisher = {IOP},
  issn = {0004-6280},
  doi = {10.1088/1538-3873/aae904},
  journal = {\pasp}
}

@article{perleyZwickyTransientFacility2020,
  title = {The {{Zwicky Transient Facility Bright Transient Survey}}. {{II}}. {{A Public Statistical Sample}} for {{Exploring Supernova Demographics}}},
  author = {Perley, Daniel A. and Fremling, Christoffer and Sollerman, Jesper and Miller, Adam A. and Dahiwale, Aishwarya S. and Sharma, Yashvi and Bellm, Eric C. and Biswas, Rahul and Brink, Thomas G. and Bruch, Rachel J. and De, Kishalay and Dekany, Richard and Drake, Andrew J. and Duev, Dmitry A. and Filippenko, Alexei V. and {Gal-Yam}, Avishay and Goobar, Ariel and Graham, Matthew J. and Graham, Melissa L. and Ho, Anna Y. Q. and Irani, Ido and Kasliwal, Mansi M. and Kim, Young-Lo and Kulkarni, S. R. and Mahabal, Ashish and Masci, Frank J. and Modak, Shaunak and Neill, James D. and Nordin, Jakob and Riddle, Reed L. and Soumagnac, Maayane T. and Strotjohann, Nora L. and Schulze, Steve and Taggart, Kirsty and Tzanidakis, Anastasios and Walters, Richard S. and Yan, Lin},
  year = 2020,
  month = nov,
  volume = {904},
  pages = {35},
  publisher = {IOP},
  issn = {0004-637X},
  doi = {10.3847/1538-4357/abbd98},
  journal = {\apj}
}

@article{perlmutterMeasurements42HighRedshift1999,
  title = {Measurements of {{\textohm}} and {{$\Lambda$}} from 42 {{High-Redshift Supernovae}}},
  author = {Perlmutter, S. and Aldering, G. and Goldhaber, G. and Knop, R. A. and Nugent, P. and Castro, P. G. and Deustua, S. and Fabbro, S. and Goobar, A. and Groom, D. E. and Hook, I. M. and Kim, A. G. and Kim, M. Y. and Lee, J. C. and Nunes, N. J. and Pain, R. and Pennypacker, C. R. and Quimby, R. and Lidman, C. and Ellis, R. S. and Irwin, M. and McMahon, R. G. and {Ruiz-Lapuente}, P. and Walton, N. and Schaefer, B. and Boyle, B. J. and Filippenko, A. V. and Matheson, T. and Fruchter, A. S. and Panagia, N. and Newberg, H. J. M. and Couch, W. J. and Project, The Supernova Cosmology},
  year = 1999,
  month = jun,
  volume = {517},
  pages = {565--586},
  publisher = {IOP},
  issn = {0004-637X},
  doi = {10.1086/307221},
  journal = {\apj}
}

@article{restCosmologicalConstraintsMeasurements2014,
  title = {Cosmological {{Constraints}} from {{Measurements}} of {{Type Ia Supernovae Discovered}} during the {{First}} 1.5 Yr of the {{Pan-STARRS1 Survey}}},
  author = {Rest, A. and Scolnic, D. and Foley, R. J. and Huber, M. E. and Chornock, R. and Narayan, G. and Tonry, J. L. and Berger, E. and Soderberg, A. M. and Stubbs, C. W. and Riess, A. and Kirshner, R. P. and Smartt, S. J. and Schlafly, E. and Rodney, S. and Botticella, M. T. and Brout, D. and Challis, P. and Czekala, I. and Drout, M. and Hudson, M. J. and Kotak, R. and Leibler, C. and Lunnan, R. and Marion, G. H. and McCrum, M. and Milisavljevic, D. and Pastorello, A. and Sanders, N. E. and Smith, K. and Stafford, E. and Thilker, D. and Valenti, S. and {Wood-Vasey}, W. M. and Zheng, Z. and Burgett, W. S. and Chambers, K. C. and Denneau, L. and Draper, P. W. and Flewelling, H. and Hodapp, K. W. and Kaiser, N. and Kudritzki, R.-P. and Magnier, E. A. and Metcalfe, N. and Price, P. A. and Sweeney, W. and Wainscoat, R. and Waters, C.},
  year = 2014,
  month = nov,
  volume = {795},
  pages = {44},
  publisher = {IOP},
  issn = {0004-637X},
  doi = {10.1088/0004-637X/795/1/44},
  journal = {\apj}
}

@article{riessObservationalEvidenceSupernovae1998,
  title = {Observational {{Evidence}} from {{Supernovae}} for an {{Accelerating Universe}} and a {{Cosmological Constant}}},
  author = {Riess, Adam G. and Filippenko, Alexei V. and Challis, Peter and Clocchiatti, Alejandro and Diercks, Alan and Garnavich, Peter M. and Gilliland, Ron L. and Hogan, Craig J. and Jha, Saurabh and Kirshner, Robert P. and Leibundgut, B. and Phillips, M. M. and Reiss, David and Schmidt, Brian P. and Schommer, Robert A. and Smith, R. Chris and Spyromilio, J. and Stubbs, Christopher and Suntzeff, Nicholas B. and Tonry, John},
  year = 1998,
  month = sep,
  volume = {116},
  pages = {1009--1038},
  publisher = {IOP},
  issn = {0004-6256},
  doi = {10.1086/300499},
  journal = {\aj}
}

@article{rigaultFullyAutomatedIntegral2019,
  title = {Fully Automated Integral Field Spectrograph Pipeline for the {{SEDMachine}}: Pysedm},
  shorttitle = {Fully Automated Integral Field Spectrograph Pipeline for the {{SEDMachine}}},
  author = {Rigault, M. and Neill, J. D. and Blagorodnova, N. and Dugas, A. and Feeney, M. and Walters, R. and Brinnel, V. and Copin, Y. and Fremling, C. and Nordin, J. and Sollerman, J.},
  year = 2019,
  month = jul,
  volume = {627},
  pages = {A115},
  publisher = {EDP},
  issn = {0004-6361},
  doi = {10.1051/0004-6361/201935344},
  journal = {\aap}
}

@article{rigaultZTFSNIa2025a,
  title = {{{ZTF SN Ia DR2}}: {{Overview}}},
  shorttitle = {{{ZTF SN Ia DR2}}},
  author = {Rigault, M. and Smith, M. and Goobar, A. and Maguire, K. and Dimitriadis, G. and Johansson, J. and Nordin, J. and Burgaz, U. and Dhawan, S. and Sollerman, J. and Regnault, N. and Kowalski, M. and Nugent, P. and Andreoni, I. and Amenouche, M. and Aubert, M. and {Barjou-Delayre}, C. and Bautista, J. and Bellm, E. and Betoule, M. and Bloom, J. S. and Carreres, B. and Chen, T. X. and Copin, Y. and Deckers, M. and {de Jaeger}, T. and Feinstein, F. and Fouchez, D. and Fremling, C. and Galbany, L. and Ginolin, M. and Graham, M. and Groom, S. L. and Harvey, L. and Kasliwal, M. M. and Kenworthy, W. D. and Kim, Y.-L. and Kuhn, D. and Kulkarni, S. R. and Lacroix, L. and Laher, R. R. and Masci, F. J. and {M{\"u}ller-Bravo}, T. E. and Miller, A. and Osman, M. and Perley, D. and Popovic, B. and Purdum, J. and Qin, Y.-J. and Racine, B. and Reusch, S. and Riddle, R. and Rosnet, P. and Rosselli, D. and Ruppin, F. and Senzel, R. and Rusholme, B. and Schweyer, T. and Terwel, J. H. and Townsend, A. and Tzanidakis, A. and Wold, A. and Yan, L.},
  year = 2025,
  month = feb,
  volume = {694},
  pages = {A1},
  publisher = {EDP},
  issn = {0004-6361},
  doi = {10.1051/0004-6361/202450388},
  journal = {\aap}
}

@article{rigaultZTFSNIa2025b,
  title = {{{ZTF SN Ia DR2}}: {{Study}} of {{Type Ia}} Supernova Light-Curve Fits},
  shorttitle = {{{ZTF SN Ia DR2}}},
  author = {Rigault, M. and Smith, M. and Regnault, N. and Kenworthy, W. D. and Maguire, K. and Goobar, A. and Dimitriadis, G. and Johansson, J. and Amenouche, M. and Aubert, M. and {Barjou-Delayre}, C. and Bellm, E. C. and Burgaz, U. and Carreres, B. and Copin, Y. and Deckers, M. and {de Jaeger}, T. and Dhawan, S. and Feinstein, F. and Fouchez, D. and Galbany, L. and Ginolin, M. and Graham, M. J. and Kim, Y.-L. and Kowalski, M. and Kuhn, D. and Kulkarni, S. R. and {M{\"u}ller-Bravo}, T. E. and Nordin, J. and Popovic, B. and Purdum, J. and Rosnet, P. and Rosselli, D. and Racine, B. and Ruppin, F. and Sollerman, J. and Terwel, J. H. and Townsend, A.},
  year = 2025,
  month = feb,
  volume = {694},
  pages = {A2},
  publisher = {EDP},
  issn = {0004-6361},
  doi = {10.1051/0004-6361/202450377},
  journal = {\aap}
}

@article{rubinLookingLambdaUnion2009,
  title = {Looking {{Beyond Lambda}} with the {{Union Supernova Compilation}}},
  author = {Rubin, D. and Linder, E. V. and Kowalski, M. and Aldering, G. and Amanullah, R. and Barbary, K. and Connolly, N. V. and Dawson, K. S. and Faccioli, L. and Fadeyev, V. and Goldhaber, G. and Goobar, A. and Hook, I. and Lidman, C. and Meyers, J. and Nobili, S. and Nugent, P. E. and Pain, R. and Perlmutter, S. and {Ruiz-Lapuente}, P. and Spadafora, A. L. and Strovink, M. and Suzuki, N. and Swift, H.},
  year = 2009,
  month = apr,
  volume = {695},
  pages = {391--403},
  publisher = {IOP},
  issn = {0004-637X},
  doi = {10.1088/0004-637X/695/1/391},
  journal = {\apj}
}

@inproceedings{rubinUnionUNITYCosmology2025,
  title = {Union {{Through UNITY}}: {{Cosmology}} with 2,000 {{SNe Using}} a {{Unified Bayesian Framework}}},
  shorttitle = {Union {{Through UNITY}}},
  booktitle = {American {{Astronomical Society Meeting Abstracts}} \#245},
  author = {Rubin, David and Aldering, Greg and Betoule, Marc and Fruchter, Andrew and Huang, Xiaosheng and Kim, Alex and Lidman, Chris and Linder, Eric and Perlmutter, Saul and {Ruiz-Lapuente}, Pilar and Suzuki, Nao},
  year = 2025,
  month = jan,
  volume = {245},
  pages = {250.02}
}

@article{schlaflyMeasuringReddeningSloan2011,
  title = {Measuring {{Reddening}} with {{Sloan Digital Sky Survey Stellar Spectra}} and {{Recalibrating SFD}}},
  author = {Schlafly, Edward F. and Finkbeiner, Douglas P.},
  year = 2011,
  month = aug,
  volume = {737},
  pages = {103},
  publisher = {IOP},
  issn = {0004-637X},
  doi = {10.1088/0004-637X/737/2/103},
  journal = {\apj}
}

@article{schmidtHighZSupernovaSearch1998,
  title = {The {{High-Z Supernova Search}}: {{Measuring Cosmic Deceleration}} and {{Global Curvature}} of the {{Universe Using Type IA Supernovae}}},
  shorttitle = {The {{High-Z Supernova Search}}},
  author = {Schmidt, Brian P. and Suntzeff, Nicholas B. and Phillips, M. M. and Schommer, Robert A. and Clocchiatti, Alejandro and Kirshner, Robert P. and Garnavich, Peter and Challis, Peter and Leibundgut, B. and Spyromilio, J. and Riess, Adam G. and Filippenko, Alexei V. and Hamuy, Mario and Smith, R. Chris and Hogan, Craig and Stubbs, Christopher and Diercks, Alan and Reiss, David and Gilliland, Ron and Tonry, John and Maza, Jos{\'e} and Dressler, A. and Walsh, J. and Ciardullo, R.},
  year = 1998,
  month = nov,
  volume = {507},
  pages = {46--63},
  publisher = {IOP},
  issn = {0004-637X},
  doi = {10.1086/306308},
  journal = {\apj}
}

@misc{spergelWideFieldInfrarRedSurvey2015,
  title = {Wide-{{Field InfrarRed Survey Telescope-Astrophysics Focused Telescope Assets WFIRST-AFTA}} 2015 {{Report}}},
  author = {Spergel, D. and Gehrels, N. and Baltay, C. and Bennett, D. and Breckinridge, J. and Donahue, M. and Dressler, A. and Gaudi, B. S. and Greene, T. and Guyon, O. and Hirata, C. and Kalirai, J. and Kasdin, N. J. and Macintosh, B. and Moos, W. and Perlmutter, S. and Postman, M. and Rauscher, B. and Rhodes, J. and Wang, Y. and Weinberg, D. and Benford, D. and Hudson, M. and Jeong, W.-S. and Mellier, Y. and Traub, W. and Yamada, T. and Capak, P. and Colbert, J. and Masters, D. and Penny, M. and Savransky, D. and Stern, D. and Zimmerman, N. and Barry, R. and Bartusek, L. and Carpenter, K. and Cheng, E. and Content, D. and Dekens, F. and Demers, R. and Grady, K. and Jackson, C. and Kuan, G. and Kruk, J. and Melton, M. and Nemati, B. and Parvin, B. and Poberezhskiy, I. and Peddie, C. and Ruffa, J. and Wallace, J. K. and Whipple, A. and Wollack, E. and Zhao, F.},
  year = 2015,
  month = mar,
  publisher = {arXiv},
  doi = {10.48550/arXiv.1503.03757}
}

@article{stetsonDAOPHOTComputerProgram1987,
  title = {{{DAOPHOT}}: {{A Computer Program}} for {{Crowded-Field Stellar Photometry}}},
  shorttitle = {{{DAOPHOT}}},
  author = {Stetson, Peter B.},
  year = 1987,
  month = mar,
  volume = {99},
  pages = {191},
  publisher = {IOP},
  issn = {0004-6280},
  doi = {10.1086/131977},
  journal = {\pasp}
}

@article{sullivanSNLS3ConstraintsDark2011,
  title = {{{SNLS3}}: {{Constraints}} on {{Dark Energy Combining}} the {{Supernova Legacy Survey Three-year Data}} with {{Other Probes}}},
  shorttitle = {{{SNLS3}}},
  author = {Sullivan, M. and Guy, J. and Conley, A. and Regnault, N. and Astier, P. and Balland, C. and Basa, S. and Carlberg, R. G. and Fouchez, D. and Hardin, D. and Hook, I. M. and Howell, D. A. and Pain, R. and {Palanque-Delabrouille}, N. and Perrett, K. M. and Pritchet, C. J. and Rich, J. and {Ruhlmann-Kleider}, V. and Balam, D. and Baumont, S. and Ellis, R. S. and Fabbro, S. and Fakhouri, H. K. and Fourmanoit, N. and {Gonz{\'a}lez-Gait{\'a}n}, S. and Graham, M. L. and Hudson, M. J. and Hsiao, E. and Kronborg, T. and Lidman, C. and Mourao, A. M. and Neill, J. D. and Perlmutter, S. and Ripoche, P. and Suzuki, N. and Walker, E. S.},
  year = 2011,
  month = aug,
  volume = {737},
  pages = {102},
  publisher = {IOP},
  issn = {0004-637X},
  doi = {10.1088/0004-637X/737/2/102},
  journal = {\apj}
}

@article{suzukiHubbleSpaceTelescope2012,
  title = {The {{Hubble Space Telescope Cluster Supernova Survey}}. {{V}}. {{Improving}} the {{Dark-energy Constraints}} above z {$>$} 1 and {{Building}} an {{Early-type-hosted Supernova Sample}}},
  author = {Suzuki, N. and Rubin, D. and Lidman, C. and Aldering, G. and Amanullah, R. and Barbary, K. and Barrientos, L. F. and Botyanszki, J. and Brodwin, M. and Connolly, N. and Dawson, K. S. and Dey, A. and Doi, M. and Donahue, M. and Deustua, S. and Eisenhardt, P. and Ellingson, E. and Faccioli, L. and Fadeyev, V. and Fakhouri, H. K. and Fruchter, A. S. and Gilbank, D. G. and Gladders, M. D. and Goldhaber, G. and Gonzalez, A. H. and Goobar, A. and Gude, A. and Hattori, T. and Hoekstra, H. and Hsiao, E. and Huang, X. and Ihara, Y. and Jee, M. J. and Johnston, D. and Kashikawa, N. and Koester, B. and Konishi, K. and Kowalski, M. and Linder, E. V. and Lubin, L. and Melbourne, J. and Meyers, J. and Morokuma, T. and Munshi, F. and Mullis, C. and Oda, T. and Panagia, N. and Perlmutter, S. and Postman, M. and Pritchard, T. and Rhodes, J. and Ripoche, P. and Rosati, P. and Schlegel, D. J. and Spadafora, A. and Stanford, S. A. and Stanishev, V. and Stern, D. and Strovink, M. and Takanashi, N. and Tokita, K. and Wagner, M. and Wang, L. and Yasuda, N. and Yee, H. K. C. and Supernova Cosmology Project, The},
  year = 2012,
  month = feb,
  volume = {746},
  pages = {85},
  publisher = {IOP},
  issn = {0004-637X},
  doi = {10.1088/0004-637X/746/1/85},
  journal = {\apj}
}

@article{taylorRevisedSALT2Surface2021,
  title = {A Revised {{SALT2}} Surface for Fitting {{Type Ia}} Supernova Light Curves},
  author = {Taylor, G. and Lidman, C. and Tucker, B. E. and Brout, D. and Hinton, S. R. and Kessler, R.},
  year = 2021,
  month = jul,
  volume = {504},
  pages = {4111--4122},
  publisher = {OUP},
  issn = {0035-8711},
  doi = {10.1093/mnras/stab962},
  journal = {\mnras}
}

@article{trippTwoparameterLuminosityCorrection1998,
  title = {A Two-Parameter Luminosity Correction for {{Type IA}} Supernovae},
  author = {Tripp, Robert},
  year = 1998,
  month = mar,
  volume = {331},
  pages = {815--820},
  publisher = {EDP},
  issn = {0004-6361},
  journal = {\aap}
}

@article{zackayProperImageSubtraction2016,
  title = {Proper {{Image Subtraction}}---{{Optimal Transient Detection}}, {{Photometry}}, and {{Hypothesis Testing}}},
  author = {Zackay, Barak and Ofek, Eran O. and {Gal-Yam}, Avishay},
  year = 2016,
  month = oct,
  volume = {830},
  pages = {27},
  publisher = {IOP},
  issn = {0004-637X},
  doi = {10.3847/0004-637X/830/1/27},
  journal = {\apj}
}

\end{document}